\documentclass[preprint]{revtex4}
\usepackage{graphicx}
\usepackage{psfrag}
\usepackage{color}
\usepackage{ulem}
\usepackage[colorlinks,citecolor=green,urlcolor=blue,bookmarks=false,hypertexnames=true]{hyperref} 
\usepackage[outdir=./]{epstopdf}
\footnotesize
\usepackage{booktabs}
\usepackage{hyperref}
\usepackage{mathtools}
\usepackage{natbib}

\begin{document}
	
	\title{Gravitational Form Factors and Mechanical Properties of the Proton : Connection Between Distributions in 2D and 3D}
	
	\author{Poonam Choudhary$^1$, Bheemsehan Gurjar$^1$, Dipankar Chakrabarti$^1$ and Asmita Mukherjee$^2$ }

	\affiliation{$^1$ Department of Physics, Indian Institute of Technology Kanpur, Kanpur 208016, India\\
	$^2$ Department of Physics, \\Indian Institute of Technology Bombay, Powai, Mumbai 400076, India}

	\begin{abstract}
The gravitational form factors which are obtained from  the matrix elements of the energy momentum tensor provide us information about internal  distributions of mass, energy,  pressure and shear. The Druck term is the least understood among all the gravitational form factor.
In a light front quark-diquark model of proton, we investigate the Druck form factor. Using Abel transformation, we evaluate the 3D distribution in the  Breit frame from the 2D light front distributions. The results are compared with other models and lattice predictions.
	\end{abstract}
	\maketitle
	%\tableofcontents
	\section{INTRODUCTION}
	\label{sec:Introduction}
	 Form factors are sources of information about the internal structure of hadron. There are several form factors that give a different kind of information about the hadron. Hadron structures are probed by electromagnetic, weak, and gravitational interactions where particles couple to the matter fields. In electromagnetic interaction photon couples to matter field and the corresponding 
	conserved electromagnetic current gives electric and magnetic charge distributions inside the hadron through form factors. The weak interaction which is mediated through $ W_{\pm}$  and $ Z $ bosons provide axial and pseudoscalar form factors of the proton. The gravitational interaction between graviton-proton provides mass, spin, and force distribution inside the proton \cite{Polyakov:2018zvc,Lorce:2018egm}. Corresponding form factors are known as gravitational form factors and are written as matrix elements of the energy-momentum tensor (EMT) \cite{Kobzarev:1962wt,Pagels:1966zza}.
	The electromagnetic and weak properties of the proton are well known but the internal mass or energy distributions, the forces on the quarks, and the angular momentum distribution inside the proton have attracted a lot of attention only recently.\\ 
	In the forward limit, the electromagnetic form factors are equivalent to electric charge and magnetic moment and the weak interaction form factors are equivalent to the axial charge and pseudoscalar coupling while gravitational interaction describes mass, spin, and D-term \cite{Polyakov:1999gs,Kivel:2000fg} in this limit. The matrix element of EMT describes the response of the nucleon to a change in the external space-time metric. The components of the energy-momentum tensor tell us how matter couples to the gravitational field. The gravitational form factors  are accessible through hard exclusive processes like deeply virtual Compton scattering as the second moments of Generalized Parton distribution functions (GPDs) \cite{Muller:1994ses,Ji:1996ek,Radyushkin:1996nd,Radyushkin:1996ru}. In \cite{Rajan:2018zzy} a connection had been established between observables
from high energy experiments and from the analysis of gravitational wave events. In the standard EMT parametrization, there are three gravitational form factors (GFFs). The GFFs $ A(Q^2)$, $J(Q^2)$, and $ D(Q^2)$ correspond to the time-time, time-space, and space-space components of the energy-momentum tensor, respectively. At zero momentum transfer, $ Q^2 =0 $ the GFFs $ A(Q^2) $ and $ J(Q^2)$ are constrained by proton mass and spin respectively. The D-term form factor is the new and most exciting one which is extracted through the spatial-spatial component of the energy-momentum tensor and encodes the information on shear forces and pressure distribution inside the proton \cite{Polyakov:2002yz}. It has been calculated in several models and theories in the literature. In \cite{Polyakov:2002yz} it was shown for the first time how GPDs can give information on the mechanical properties of the proton in a DVCS process, as they are  extracted from the beam charge asymmetry in deeply virtual Compton scattering. While in \cite{Burkert:2018bqq,Burkert:2021ith} the JLab group reported the first determination of the pressure and shear forces on quarks inside the proton from experimental data on deeply virtual Compton scattering. The EMT form factors of the nucleon have been investigated in various approaches,
	for example, in lattice QCD \cite{Gockeler:2003jfa,Hagler:2003jd,Shanahan:2018nnv} in chiral perturbation theory \cite{Dorati:2007bk,Chen:2001pva,Belitsky:2002jp,Diehl}, in the chiral quark-soliton model \cite{Goeke:2007fp,Kim:2021jjf} as well as in the Skyrme model \cite{Cebulla:2007ei,Kim:2012ts}. 
	
	The pressure, shear and energy distributions are usually defined in terms of the static EM tensor in the Breit frame. In relativistic field theory, one cannot localize a particle within a Compton wavelength, in other words, the three dimensional distributions defined in the Breit frame are subject to relativistic corrections. Alternatively, for a relativistic system, one can define them in the so-called infinite momentum frame, or light front quantization, where such relativistic effects are already incorporated. In \cite{Lorce:2018egm} the mechanical properties like pressure, shear and energy distributions of a nucleon in two dimensions were introduced in the light front formalism, also later were discussed in \cite{Freese:2021czn}. %[Freese and Miller. 2102.01683].
	While in some models one can easily calculate the 3D distributions, in some cases, for example in light-front wavefunction approach it is easier to calculate the 2D distributions. In \cite{Panteleeva:2021iip} it was shown that the 2D and 3D distributions can be connected through Abel transformation, which would make the intuitive understanding of these distributions more clear, particularly for relativistic systems like a nucleon. In a previous  work, \cite{Chakrabarti:2020kdc} the GFFs and the two-dimensional pressure, shear and energy distributions were investigated in the spectator type model motivated by ADs/QCD. %The matrix elements of the total EMT in one-particle states define the EMT form %factors which are Lorentz scalars. Due to the EMT conservation, the form factors %of the total EMT are renormalization scale-invariant. The separate quark and gluon %EMT operators have not conserved additional form factors that appear in the %decompositions of their matrix elements, and all individual quark and gluon form %factors acquire scale $ \mu $- and renormalization scheme-dependence. There has %been several work on quark EMT form factors \cite{Chakrabarti:2020kdc} %cite{Anikin:2019kwi} \cite{Goeke:2007fp}.
	In this work, we use the same model to obtain pressure, shear and energy distributions of the nucleon in 3D using invertible Abel transformation. The outline of the present work is as follows: in \ref{LFQDQmodel} we briefly review the light front formalism based on the light-front quark diquark model. In \ref{EMTFFs} we illustrate the definition of the gravitational form factors as matrix elements of the energy-momentum tensor and define the form factors in LFQDQ. Then in \ref{GFFLFQDQ} we focus on the extraction of $ D(Q^2)$ using two different approaches. And in \ref{modelresults} we show the 3D BF distributions which are the Abel image of the 2D light-front distributions by doing the inverse Abel transformation. And finally, in \ref{conclusion} we present the summary and conclusion.
	%%%%%%%%%%%%%%%
	\section{LIGHT-FRONT QUARK-DIQUARK MODEL}\label{LFQDQmodel}
	In this model,  the incoming photon, carrying a high momentum, interacts with one of the valence quarks inside the nucleon, and the other two valence quarks form a spectator diquark state of spin-0 (scalar diquark). Therefore the nucleon state $|P,S\rangle$ having momentum $P$ and spin $S$, can be represented as a two-particle Fock-state. In this article we consider the quark-scalar diquark model proposed in\cite{Gutsche:2013zia}. We use the  light-cone convention $x^{\pm}=x^{0}\pm x^{3}$, and choose a frame where the transverse momentum of the proton vanishes, i.e. $P\equiv\left(P^{+},\frac{M^{2}}{P^{+}},\mathbf{0}_{\perp}\right)$, while the momentum of the quark and the diquark are $p\equiv\left(xP^{+},\frac{p^{2}+|\mathbf{p}_{\perp}^{2}|}{xP^{+}},\mathbf{p}_{\perp}\right)$ and $P_{X}\equiv\left((1-x)P^{+},P_{X}^{-},-\mathbf{p}_{\perp}\right)$ respectively, where $x=p^{+}/P^{+}$ is the longitudinal momentum fraction of the active quark. The two particle Fock-state expansion for the state with helicity $\pm \frac{1}{2}$ is given by
	\begin{eqnarray} \label{protonstate}
		|P ; \pm\rangle&=& \sum_{q} \int \frac{d x d^{2} \mathbf{p}_{\perp}}{2(2 \pi)^{3} \sqrt{x(1-x)}}\nonumber \\  
		&&\times\left[\psi_{+}^{q \pm}\left(x, \mathbf{p}_{\perp}\right)\left|+\frac{1}{2}, 0 ; x P^{+}, \mathbf{p}_{\perp}\right\rangle+\psi_{-}^{q \pm}\left(x, \mathbf{p}_{\perp}\right)\left|-\frac{1}{2}, 0 ; x P^{+}, \mathbf{p}_{\perp}\right\rangle\right], 
	\end{eqnarray}
	where $|\lambda_{q},\lambda_{s};xP^{+},\mathbf{p}_{\perp}\rangle$ represents the two particle state with a quark having spin $\lambda_{q}=\pm \frac{1}{2}$, momentum $p$ and a scalar spectator diquark with spin $\lambda_{S}=0$. The two particle states are normalized as
	\begin{equation} \label{fockstate}
		\left\langle\lambda_{q}^{\prime}, \lambda_{s}^{\prime} ; x^{\prime} P^{+}, \mathbf{p}_{\perp}^{\prime} \mid \lambda_{q}, \lambda_{s} ; x P^{+}, \mathbf{p}_{\perp}\right\rangle=\prod_{i=1}^{2} 16 \pi^{3} p_{i}^{+} \delta\left(p_{i}^{\prime+}-p_{i}^{+}\right) \delta^{2}\left(\mathbf{p}_{\perp i}^{\prime}-\mathbf{p}_{\perp i}\right) \delta_{\lambda_{i}^{\prime} \lambda_{i}} .
	\end{equation}	
	Here $\psi_{\lambda_{q}}^{q \lambda_{N}}$ are the light-front wave functions with nucleon helicities $\lambda_{N}=\pm$. % We adopt the generic ansatz for the quark-diquark model of the valence Fock state of the nucleon
	The LFWFs  are given by\cite{Gutsche:2013zia}.
	\begin{eqnarray}\label{LFWF}
		&\psi_{+}^{q+}\left(x, \mathbf{p}_{\perp}\right)&=\varphi^{q(1)}\left(x, \mathbf{p}_{\perp}\right) \nonumber \\
		&\psi_{-}^{q+}\left(x, \mathbf{p}_{\perp}\right)&=-\frac{p^{1}+i p^{2}}{x M} \varphi^{q(2)}\left(x, \mathbf{p}_{\perp}\right) \nonumber  \\
		&\psi_{+}^{q-}\left(x, \mathbf{p}_{\perp}\right)&=\frac{p^{1}-i p^{2}}{x M} \varphi^{q(2)}\left(x, \mathbf{p}_{\perp}\right) \nonumber \\
		&\psi_{-}^{q-}\left(x, \mathbf{p}_{\perp}\right)&=\varphi^{q(1)}\left(x, \mathbf{p}_{\perp}\right)
	\end{eqnarray}
	where $\varphi_{q}^{(1)}(x,\mathbf{p}_{\perp})$ and $\varphi_{q}^{(2)}(x,\mathbf{p}_{\perp})$ are the wave functions predicted by the soft-wall AdS/QCD and can be written as
	\begin{eqnarray}  \label{phiAdSQCD}
		\varphi^{q(i)}\left(x, \mathbf{p}_{\perp}\right)=N_{q}^{(i)} \frac{4 \pi}{\kappa} \sqrt{\frac{\log (1 / x)}{1-x}} x^{a_{q}^{(i)}}(1-x)^{b_{q}^{(i)}} \exp \left[-\frac{\mathbf{p}_{\perp}^{2}}{2 \kappa^{2}} \frac{\log (1 / x)}{(1-x)^{2}}\right];
	\end{eqnarray} % \label{Eq4}
	where $\kappa=0.4$ GeV is the AdS/QCD scale parameter and the quarks are assumed to be massless \cite{Brodsky:2006ha}. The values of the model parameters $a_{q}^{i}$ and $b_{q}^{i}$ and the normalization constants $N_{q}^{i}$ at an initial scale $\mu_{0}^{2}=0.32$ GeV$^2$ were fixed by fitting the nucleon electromagnetic form factors and can be found in ref. \cite{Chakrabarti:2015ama}. The wave functions can be reduced to the form predicted by AdS/QCD for  $a_{q}^{i}=b_{q}^{i}=0$ \cite{deTeramond:2011aml}.
	%%%%%%%%%%%%%%%%
	\section{RELATIONS BETWEEN 2D LIGHT FRONT DISTRIBUTIONS AND 3D BREIT FRAME DISTRIBUTIONS}\label{EMTFFs}
	%%%%%%%%%%%%%%%%%%%%%%%%%%
	As discussed in the Introduction, in case of  nucleons there are three independent EMT form factors \cite{Ji:1996ek,Ji:2012vj,Harindranath:2013goa,Pagels:1966zza,Kobzarev:1962wt}	
	\begin{eqnarray}\label{QCDEMT}
		\left\langle p^{\prime}\left|\hat{\Theta}_{\mathrm{QCD}}^{\mu \nu}(0)\right| p\right\rangle=& \bar{u}\left(p^{\prime}\right)\left[A(t) \frac{P^{\mu} P^{\nu}}{M}+J(t) \frac{i P^{\{\mu} \sigma^{\nu\}{\alpha}}  \Delta_{\alpha}}{M}\right.\nonumber \\
		&\left.+\frac{D(t)}{4 M}\left(\Delta^{\mu} \Delta^{\nu}-\eta^{\mu \nu} \Delta^{2}\right)\right] u(p),
	\end{eqnarray}	
	where $\hat{\Theta}_{\mathrm{QCD}}^{\mu \nu}(x)$ is the symmetric EMT operator of QCD. $P=(p+p^{\prime})/2, \Delta=p^{\prime}-p, t=\Delta^{2}$, and the symmetrization operator is defined as $X_{\{\mu} Y_{\nu\}}=\frac{1}{2}(X_{\mu}Y_{\nu}+X_{\nu}Y_{\mu})$. At zero momentum transfer the values of these nucleon EMT  form factors (FFs) provide us with the three basic characteristics of the nucleons: the mass M, spin $J=1/2$, and the D-term (also known as Druck term) D(0). The mass and the spin of the nucleons are well-observed quantities, the third mechanical characteristic, the D-term is more subtle term, as it is related to the distribution of the internal forces inside the nucleons \cite{Polyakov:2002yz}. For the D-term, the first experimental data are available for the nucleons \cite{Burkert:2018bqq,Kumericki:2019ddg}.
	Recently in Refs.\cite{Lorce:2018egm,Freese:2021czn} the 2D light front pressure and shear force distributions were obtained in terms of the Druck form factor D(t) as
	\begin{eqnarray}\label{FTDterm}
		\tilde{D}\left(x_{\perp}\right) &=\frac{1}{4 P^{+}} \int \frac{d^{2} \Delta_{\perp}}{(2 \pi)^{2}} D\left(-\Delta_{\perp}^{2}\right) e^{-i \Delta_{\perp} \cdot x_{\perp}},
	\end{eqnarray}
	\begin{eqnarray} \label{2Dpressure}
		p^{(2 D)}\left(x_{\perp}\right) &=\frac{1}{2 x_{\perp}} \frac{d}{d x_{\perp}}\left(x_{\perp} \frac{d}{d x_{\perp}} \tilde{D}\left(x_{\perp}\right)\right), 
	\end{eqnarray}
	\begin{eqnarray}\label{2DShear}
		s^{(2 D)}\left(x_{\perp}\right) &=-x_{\perp} \frac{d}{d x_{\perp}}\left(\frac{1}{x_{\perp}} \frac{d}{d x_{\perp}} \tilde{D}\left(x_{\perp}\right)\right),
	\end{eqnarray}
	where $x_{\perp}$ is the 2D position vector in the transverse plane. Since the 2D and 3D force distributions are expressed in terms of the same Druck term form factor $D(t)$, those distributions can be related to each other. To establish the relations between 2D and 3D distributions, it is convenient to redefine the 2D pressure $\mathcal{P}(x_{\perp})$ and shear force distributions $\mathcal{S}(x_{\perp})$  by multiplying   Eqs.(\ref{2Dpressure}) and (\ref{2DShear})  with the Lorentz factor $\frac{P^{+}}{2M}$ \cite{Lorce:2018egm} i.e.,
	\begin{eqnarray}
		\mathcal{S}(x_{\perp})&=&\frac{P^{+}}{2M}s^{(2D)}(x_{\perp}), \hspace{1cm} \mathcal{P}(x_{\perp})=\frac{P^{+}}{2M}p^{(2D)}(x_{\perp}).
	\end{eqnarray}
	By using the Abel transformation \cite{Polyakov:2007rv, Moiseeva:2008qd}, these 2D LF distributions can be  related to  the 3D distributions in the Breit frame as \cite{Panteleeva:2021iip}
	\begin{eqnarray}
		\frac{\mathcal{S}(x_\perp)}{x_\perp^2}=\int_{x_{\perp}}^{\infty}\frac{dr}{r}s(r)\frac{1}{\sqrt{r^{2}-x_{\perp}^{2}}}, \label{s(x)} \\
		\frac{1}{2}\mathcal{S}(x_{\perp})+\mathcal{P}(x_{\perp})=\int_{x_{\perp}}^{\infty}\frac{dr}{r}s(r){\sqrt{r^{2}-x_{\perp}^{2}}}.\label{s(x)p(x)}
	\end{eqnarray}
		From Eq.(\ref{s(x)}), one can see that the function $\mathcal{S}(x_{\perp})/x_{\perp}^{2}$ is the Abel image of $s(r)$.
	The Breit frame distributions of the elastic pressure $p(r)$ and shear force $s(r)$ in 3D are  obtained in terms of Druck form factor $D(t)$ (see Ref. \cite{Polyakov:2018zvc,Polyakov:2002yz}) as
	\begin{eqnarray} %\label{3D shear-pressure}
		p(r)=\frac{1}{6M}\frac{1}{r^2}\frac{d}{dr}r^{2}\frac{d}{dr}\tilde{D}(r), \hspace*{1cm} s(r)=- \frac{1}{4M}r\frac{d}{dr}\frac{1}{r}\frac{d}{dr}\tilde{D}(r),
	\end{eqnarray}
	where $\tilde{D}(r)$ is the 3D Fourier transform
	of the Druck-term, i.e.,
	\begin{eqnarray}
		\tilde{D}(r)=\int \frac{d^{3}\mathbf{\Delta}}{(2\pi)^{3}}e^{-i\mathbf{\Delta}.r}D(-\mathbf{\Delta}^{2}).
	\end{eqnarray}
	The relations in Eqs.((\ref{s(x)}),(\ref{s(x)p(x)})) have the form of invertible Abel transformation \cite{Polyakov:2007rv,Moiseeva:2008qd}. 
	%The Eq.(\ref{s(x)p(x)}) can be obtained from analogous equations in Ref. \cite{Lorce:2018egm} by change of integration variable. 
	The inverse Abel transformation (3D Breit frame distribution in terms of the 2D light-front frame distributions) of the Eqs.((\ref{s(x)}),(\ref{s(x)p(x)})) can be obtained as \cite{Panteleeva:2021iip}
	\begin{eqnarray}
		s(r) &=&-\frac{2}{\pi} r^{2} \int_{r}^{\infty} d x_{\perp} \frac{d}{d x_{\perp}}\left(\frac{\mathcal{S}\left(x_{\perp}\right)}{x_{\perp}^{2}}\right) \frac{1}{\sqrt{x_{\perp}^{2}-r^{2}}}  \label{s(r)}\\
		\frac{2}{3} s(r)+p(r) &=&\frac{4}{\pi} \int_{r}^{\infty} \frac{d x_{\perp}}{x_{\perp}} \mathcal{S}\left(x_{\perp}\right) \frac{1}{\sqrt{x_{\perp}^{2}-r^{2}}} \label{s(r)p(r)}.
	\end{eqnarray}
  Eq.(\ref{s(r)p(r)}) implies that the normal force distribution in 3D i.e., [$\frac{2}{3}s(r)+p(r)$] is the Abel image of the light front shear force distribution $\mathcal{S}(x_{\perp})$ multiplied by $\frac{4}{\pi}$. Similarly the 2D distributions for the mass/energy $\mathcal{E}^{(2D)}(x_{\perp})$ and angular momentum $\rho_{J}^{(2D)}(x_{\perp})$ are obtained by using the 2D inverse Fourier transforms of GFFs $A(t)$ and $J(t)$, respectively \cite{Lorce:2017wkb,Lorce:2018egm,Panteleeva:2021iip,Freese:2021czn,Kim:2021jjf}:
	\begin{eqnarray}\label{2Denergy-angular momentum}
		\mathcal{E}^{(2D)}(x_{\perp})=P^{+}\tilde{A}(x_{\perp}), \hspace*{1cm} \rho_{J}^{(2D)}(x_{\perp})=-\frac{1}{2}x_{\perp}\frac{d}{dx}\tilde{J}(x_{\perp})
	\end{eqnarray}
	where $J(x_{\perp})$ is the angular momentum distribution in the 2D LF frame.   $\tilde{A}(x_{\perp})$ and $\tilde{J}(x_{\perp})$ are the 2D inverse Fourier transform of the corresponding GFFs, i.e.
	\begin{equation}
		\tilde{F}(x_{\perp})=\int \frac{d^{2}\Delta}{(2\pi)^{2}}e^{-i\Delta_{\perp}.x_{\perp}}F(-\Delta_{\perp}^{2}).
	\end{equation}
	Here $x_{\perp}$ and $\Delta_{\perp}$ are respectively the position and momentum vectors in the 2D plane transverse to the propagation direction of the nucleon.  
	The mass distribution can be redefined by multiplying the Lorentz factor as \cite{Kim:2021jjf}
	\begin{eqnarray}\label{2DMass}
		\mathcal{E}(x_{\perp})=\frac{M}{P^{+}}\mathcal{E}^{(2D)}(x_{\perp}), \hspace*{0.5cm} or \hspace*{0.5cm} \mathcal{E}(x_{\perp})=M \tilde{A}(x_{\perp})
	\end{eqnarray}
	Similarly,  by using the inverse Abel transformation of Eq.(\ref{2Denergy-angular momentum}) one can find the 3D Breit Frame distributions corresponding to the 2D mass and angular momentum distributions as,
	\begin{eqnarray}\label{3Dmass}
		\epsilon(r)=-\frac{1}{\pi}\int_{r}^{\infty}\frac{dx_{\perp}}{x_{\perp}}\left(\mathcal{E}(x_{\perp})\right)\frac{1}{\sqrt{x_{\perp}^{2}-r^{2}}},
	\end{eqnarray}
	and
	\begin{eqnarray}\label{3DAngularMom}
		\rho_{J}(r)=-\frac{2}{\pi}r^{2}\int_{r}^{\infty}dx_{\perp}\frac{d}{dx_{\perp}}\left(\frac{\rho_{J}(x_{\perp})}{3x_{\perp}^{2}}\right)\frac{1}{\sqrt{x_{\perp}^{2}-r^{2}}}.
	\end{eqnarray}
	%The Abel transform shows the same physical meaning from the 2D distributions to the 3D ones. This means that the 2D distributions $\mathcal{E}(x_{\perp})$ and $\rho_{J}^{(2D)}(x_{\perp})$ explains that how the mass and the angular momentum of the nucleons are distributrd in the transverse plane.
	After integrating $\mathcal{E}(x_{\perp})$ and $\rho_{J}^{(2D)}(x_{\perp})$ over $x_{\perp}$ one can get the mass and spin of the proton as
	\begin{eqnarray}
		\int d^{2}x_{\perp}\mathcal{E}(x_{\perp})=MA(0), \hspace*{1cm}{\rm and}~~ \int d^{2}x_{\perp}\rho_{J}^{(2D)}(x_{\perp})=J(0)
	\end{eqnarray}
	where the form factors are normalized as $A(0)=1$ and $J(0)=1/2$. %are the normalized form factors.
	%%%%%%%%%%%%%%%%%%%%%%%%%%%%%%%%%%%%%%%%%%%%%%%%%%%%%%%%%%%%%%%%%%%%%%%%%%%%%%%%%%%%%%%%%%%%%%%%%%%%%%%%%%
	\section{Extraction of GFFs In LFQDQ model}\label{GFFLFQDQ}
	%%%%%%%%%%%%%%%%%%%%%%%%%%%%%%
	The Form factors $A^{u+d}(Q^{2}),B^{u+d}(Q^{2})$ and $D^{u+d}(Q^{2})$ in the LFQDQ model can be parametrized in terms of structure integrals as \cite{Chakrabarti:2020kdc,Chakrabarti:2015lba}
	\begin{eqnarray}
		A^{u+d}(Q^{2})=\mathcal{I}^{u+d}_{1}(Q^2), \hspace*{1cm} B^{u+d}(Q^{2})=\mathcal{I}^{u+d}_{2}(Q^2)
	\end{eqnarray}
	and,
	\begin{eqnarray}\label{DFF}
		D^{u+d}(Q^{2})=-\frac{1}{Q^{2}}\left[2M^{2}{\mathcal{I}}^{u+d}_{1}(Q^2)-Q^{2}{\mathcal{I}}^{u+d}_{2}(Q^2)-{\mathcal{I}}^{u+d}_{3}(Q^2)\right],
	\end{eqnarray}
	where $\mathcal{I}_i^{u+d}=\mathcal{I}_i^u +\mathcal{I}_i^d $. 
	The explicit expressions of the structure integrals $\mathcal{I}_{i}^{q}(Q^{2})$ are given by \cite{Chakrabarti:2020kdc}
	\begin{equation} \label{I1q}
		\begin{aligned}
			\mathcal{I}_{1}^{q}\left(Q^{2}\right)=\int d x x\left[N_{1}^{q 2} x^{2 a_{1}^{q}}(1-x)^{2 b_{1}^{q}+1}+N_{2}^{q 2} x^{2 a_{2}^{q}-2}(1-x)^{2 b_{2}^{q}+3} \frac{1}{M^{2}}\left(\frac{k^{2}}{\log (1 / x)}-\frac{Q^{2}}{4}\right)\right]\\
			\exp \left[-\frac{\log (1 / x)}{k^{2}} \frac{Q^{2}}{4}\right],
		\end{aligned}
	\end{equation}
	\begin{eqnarray} \label{I2q}
		\mathcal{I}_{2}^{q}\left(Q^{2}\right)=2 \int d x N_{1}^{q} N_{2}^{q} x^{a_{1}^{q}+a_{2}^{q}}(1-x)^{b_{1}^{q}+b_{2}^{q}+2} \exp \left[-\frac{\log (1 / x)}{k^{2}} \frac{Q^{2}}{4}\right],
	\end{eqnarray}
	\begin{equation}  \label{I3q}
		\begin{aligned}
			\mathcal{I}_{3}^{q}\left(Q^{2}\right)=2 \int d x N_{1}^{q} N_{2}^{q} x^{a_{1}^{q}+a_{2}^{q}-2}(1-x)^{b_{1}^{q}+b_{2}^{q}+2}\left[\frac{4(1-x)^{2} x^{2}}{\log (1 / x)}+Q^{2}(1-x)^{2}-4 m^{2}\right] \\
			\exp \left[-\frac{\log (1 / x)}{k^{2}} \frac{Q^{2}}{4}\right].
		\end{aligned}
	\end{equation}
%	The parameters in Eqns.(\ref{I1q}-\ref{I3q}) are given in Ref \cite{}.
	
	The complete analytic expression for the D-term as given  above in  Eq.(\ref{DFF}) along with Eqs.((\ref{I1q}), (\ref{I2q}) and (\ref{I3q}))  is found to be very lengthy and not so intuitive. It turns out that the form factor $D(Q^{2}) \equiv D^{u+d}(Q^2)$ can be described by the multipole  function as \cite{Chakrabarti:2021mfd},
	\begin{eqnarray}\label{Dfit}
		D(Q^{2})=\frac{a}{(1+b  Q^{2})^{c}},
	\end{eqnarray}
	where the parameters $a$, $b$ and $c$ are given in the Table \ref{table1} at  the initial  scale as well as at a higher scale.
	\begin{table}[ht]
		\centering
		\begin{tabular}[t]{lccccc}
			\hline \hline
			Parameters~~ & ~~$a$~~ & ~~$b$~~ & ~~$c$~~ & \\
			\hline
			~~~$D_{0}^{\mathrm{fit}} $ &~$-18.8359$~ & ~$2.2823$~ &  ~$2.7951$~&\\
			~~~$D_{1}^{\mathrm{fit}} $ &~$-1.521$~ &~$0.531$~ & ~$3.026$~ & \\
			\hline 	\hline 
		\end{tabular}
		\caption{Fitted parameters for the fitted function $ D_{fit}(Q^2)$ Eq.(\ref{Dfit}) form factor. Here $ D_{0}^{\mathrm{fit}} $ represents the form factor at initial model scale while $D_{1}^{\mathrm{fit}}$ show the evolved form factor from initial scale $ \mu_{0}^2=0.32$ GeV$^2$ to $\mu^2=4$ GeV$^2$ in LFQDQ model.}
		\label{table1}
	\end{table}
%	By putting these Eqs.(\ref{I1q}), (\ref{I2q}) and (\ref{I3q}) into Eq.(\ref{DFF}) one can find the analytical expression for the D-term FF. 
 The comparison of the GFFs at $Q^{2}=0$ with the various phenomenological models, lattice QCD, and existing experimental data for the $D(0)$
	% is given in the Ref.\cite{Chakrabarti:2020kdc}. 
	 and  the validity for those GFFs are discussed in the reference \cite{Chakrabarti:2020kdc}. In Table \ref{table1} fitted model parameters in the first row are extracted at the initial scale $\mu_0^2=0.32~ GeV^2$, whereas the fitted parameters in the second row  correspond to the scale evolution from initial scale $\mu^{2}=0.32$ GeV$^{2}$ to the final scale $\mu^{2}=4$ GeV$^{2}$. For the evolution scheme we adopt the Dokshitzer-Gribov-Lipatov-Altarelli-Parisi (DGLAP) equations \cite{Dokshitzer:1977sg,Gribov:1972ri,Altarelli:1977zs} of QCD with next-to-next-to-leading order (NNLO) of the scale evolution. We have used the higher-order perturbative parton evolution toolkit (HOPPET) \cite{Salam:2008qg} to perform the scale evolution numerically.
	%Explicitly, during the evolution we evolve the GPDs from the initial model scale ($\mu^{2}=0.32$ GeV$^{2}$) to the relevant lattice scale ($\mu^{2}=4$ GeV$^{2}$) using the higher order perturbative parton evolution toolkit (HOPPET) \cite{Salam:2008qg}.
	We find that the QCD evolution of the GFFs $A^{u+d}(Q^{2})$, $B^{u+d}(Q^{2})$ are consistent with the lattice QCD results \cite{Chakrabarti:2020kdc,Chakrabarti:2021mfd}. Also the qualitative behavior of our D-term  is %compatiable 
	comparable with the lattice QCD \cite{LHPC:2007blg} and the experimental data from JLab \cite{Burkert:2018bqq} as well as other theoretical predictions from the KM15 global fit \cite{Kumericki:2015lhb}, dispersion relation \cite{Pasquini:2014vua}, $\chi$QSM \cite{Goeke:2007fp}, Skyrme model \cite{Cebulla:2007ei}, and bag model \cite{Ji:1997gm}.
	
\begin{figure}
		\includegraphics[scale=0.38]{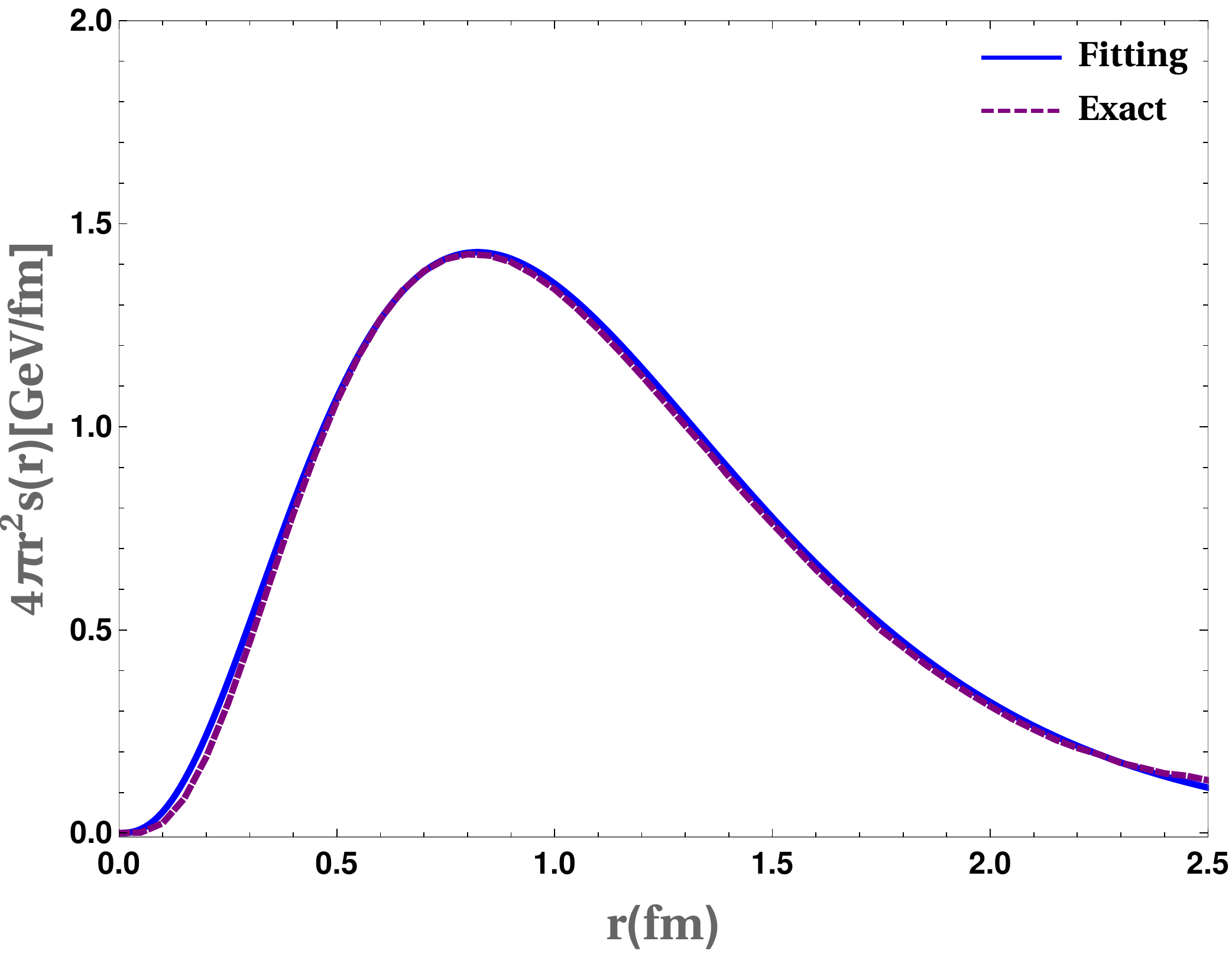}
		\includegraphics[scale=0.38]{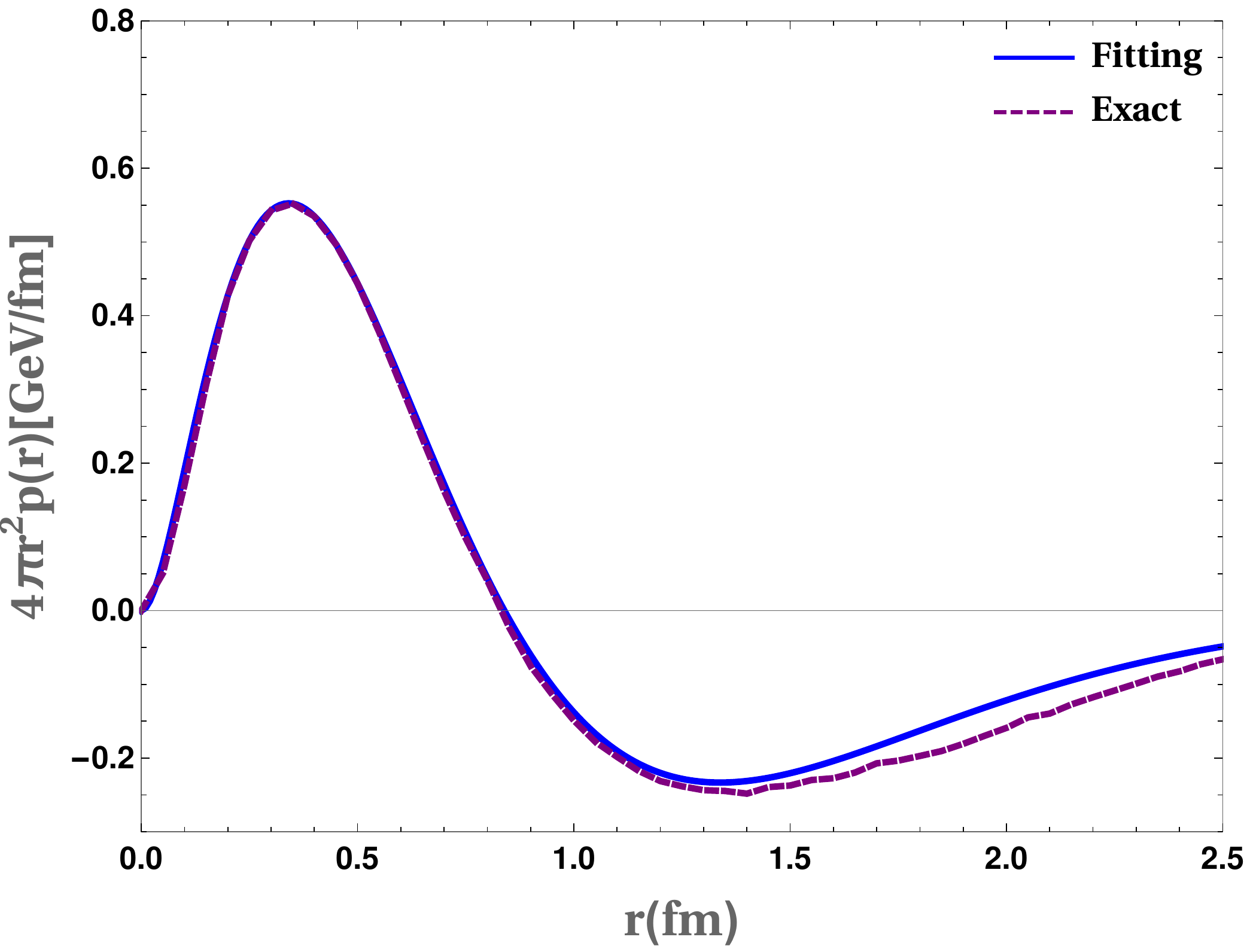}
		\includegraphics[scale=0.38]{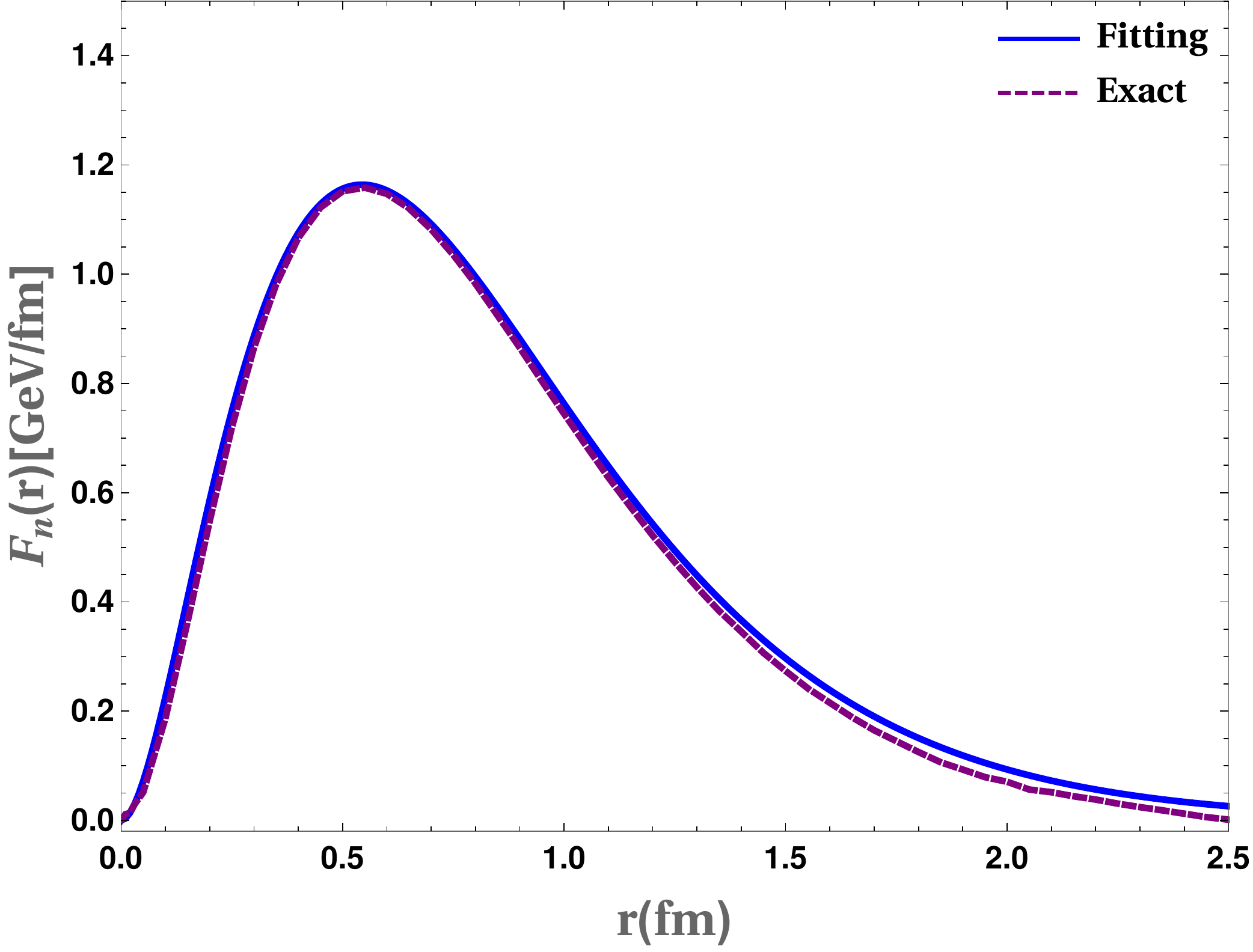}
		\includegraphics[scale=0.38]{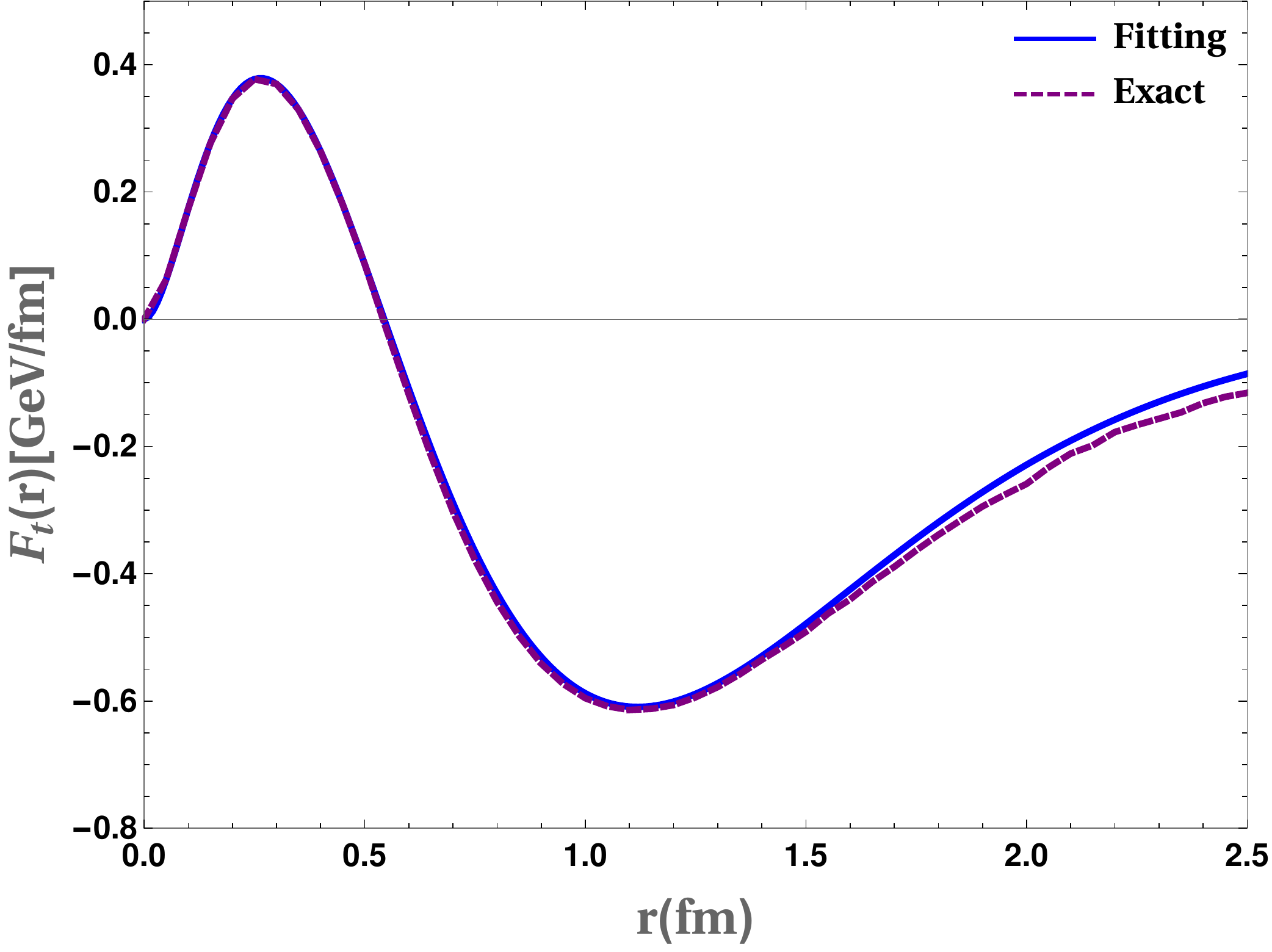}
		\caption{The solid-blue curve in the upper left panel shows the 3D shear force distribution in the Breit frame  for the LFQDQ model (using the inverse Abel transformation) at the initial scale for the fitted $D(Q^{2})$ term while the purple-dashed curve is for the model data by using the Vegas. While the upper right figure draws the 3D pressure distributions. Similarly, the lower-left panel draws the 3D normal force distribution and the lower right panel draws the 3D tangential force distribution, respectively.}
		\label{3DdistributionsInitialscale}
	\end{figure}

	We have checked the accuracy of our fitting techniques at the initial scale using the multidimensional Monte Carlo integration program  Vegas \cite{Lepage:1977sw,Lepage:2020tgj}.
	%Though we can check the accuracy of our fitting techniques at the initial scale using Monte Carlo integration like Vegas. Vegas technique \cite{Lepage:1977sw,Lepage:2020tgj} describes an algorithm for multidimensional integration, which is an iterative and adaptive Monte Carlo scheme. 
	In Fig. \ref{3DdistributionsInitialscale} we show the 3D distributions which are computed with the fitted D-term form factor given in Eq.(\ref{Dfit}) and the exact model calculations (using Vegas) at the initial scale. From Fig \ref{3DdistributionsInitialscale} one can see %that at small momentum transfer (small $Q^{2})$ the fitted and exact model results are exactly overlapping each other, while for large $Q^{2}$ those distributions are slightly differ.%
	that near the region of small spatial distance from the center of nucleon the fitted and exact model results are exactly overlapping, while for the large value of $ r $, the distributions are slightly different in two different methods. The multipole fitting function describes  the exact results  very accurately for small $r$, 
	% than large $r$, 
	but the discrepancies at large $r$ are negligibly small.  It allows us to use the multipole fitting function in place of exact expression for evaluation of different distributions using Abel transformation.
	
	\section{DISTRIBUTIONS in three dimensions }\label{modelresults}
\begin{figure}
		\includegraphics[scale=0.38]{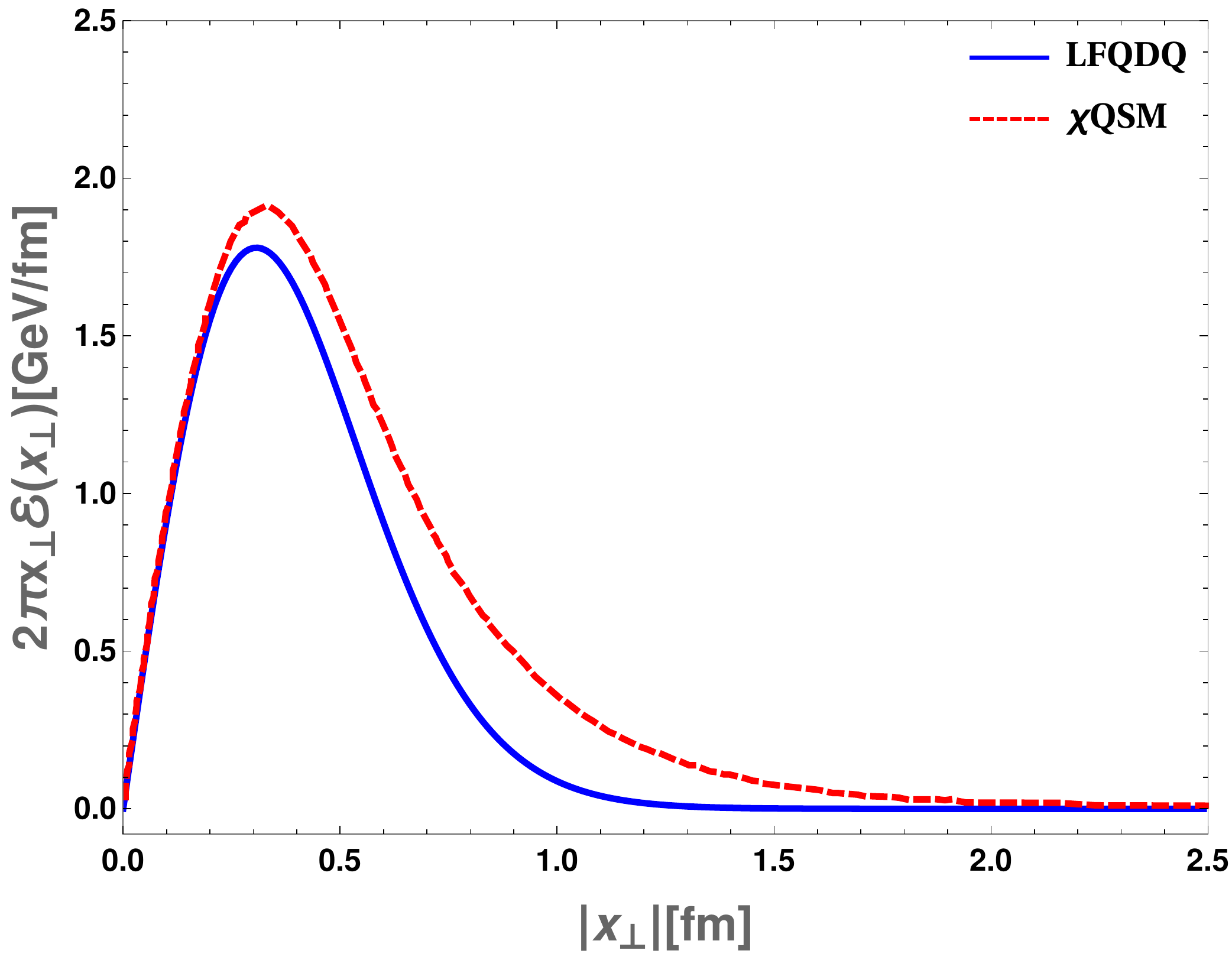}
		\includegraphics[scale=0.38]{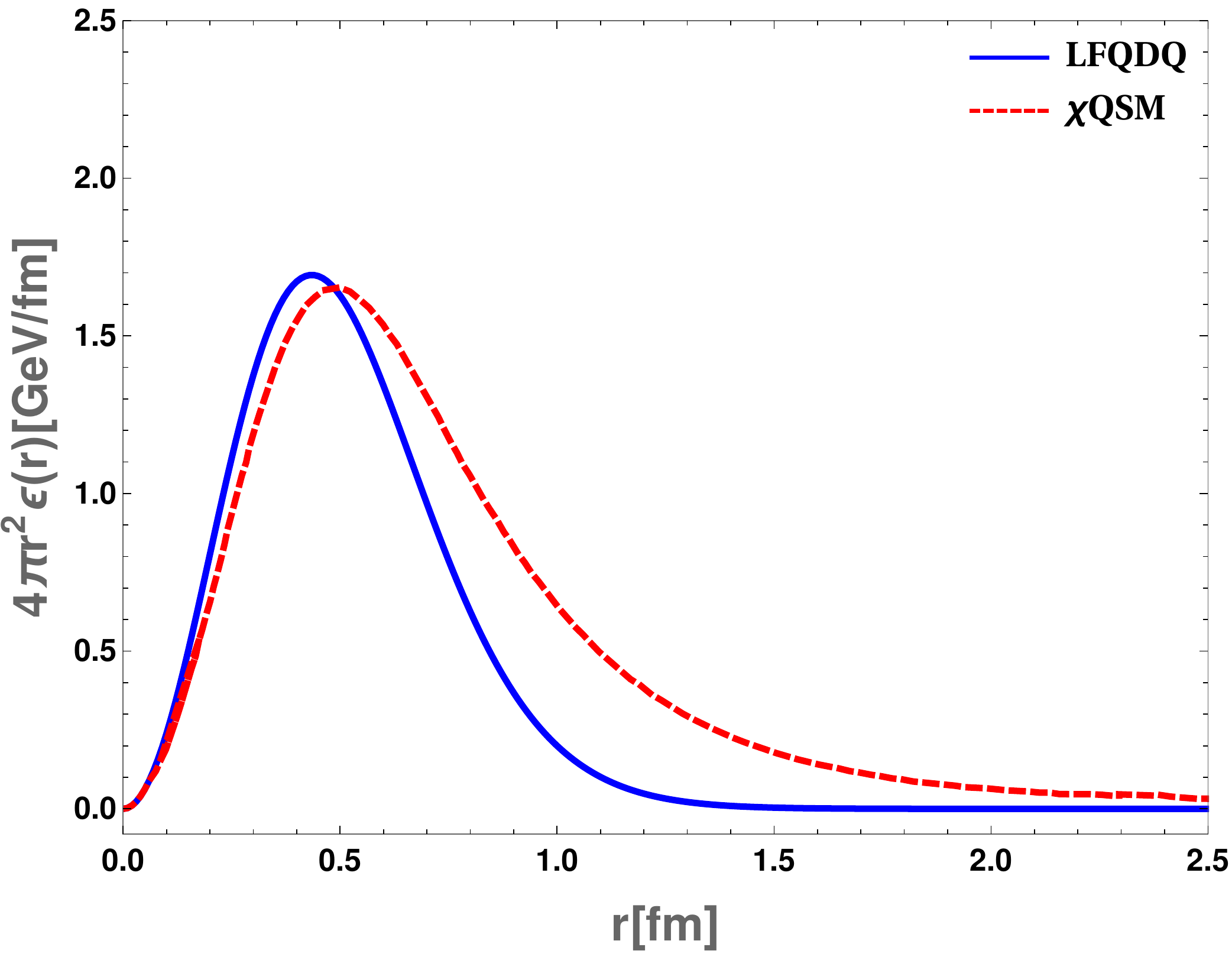}
		\caption{The solid blue curve in the left panel shows the 2D mass distribution in the LF-frame for the LFQDQ model while the red-dashed curve is the 2D mass distribution for the $\chi$QSM model \cite{Kim:2021jjf}. Similarly, the solid blue curve in the right panel shows the 3D mass distribution in the BF for the LFQDQ (after doing the inverse Abel transformation) whereas the red-dashed curve is the 3D mass distribution for $\chi$QSM model \cite{Goeke:2007fp}. Our model predictions are at evolved scale $\mu^{2}$=4 GeV$^{2}$.}
		\label{massdistributions}
	\end{figure}

	In this section, we present the results for the 3D distributions in the Breit frame (BF). The 3D BF EMT distributions are derived from the 2D LF EMT distributions by using the inverse Abel transformation\cite{Polyakov:2007rv,Moiseeva:2008qd}. \\
	In the left and right panels of Fig. \ref{massdistributions}, we compare the model results for the energy(mass) distributions  with the Chiral quark-soliton model($\chi$QSM) \cite{Goeke:2007fp,Kim:2021jjf} for the 2D and 3D mass distributions in the LF (Drell-Yan) and BF, respectively. By using  Eq.(\ref{2DMass}) we compute the 2D momentum distributions and then the 3D momentum distributions are calculated by the inverse Abel transformation defined in Eq.(\ref{3Dmass}). The values of the mass distribution in the LF-frame and BF at the center of the nucleon are found to be $\mathcal{E}(0)$=1.54 GeV/$fm^{2}$ and $\epsilon (0)$=2.02 GeV/$fm^{3}$ respectively. In Fig. \ref{massdistributions}, we have shown the  2D and 3D mass distributions weighted by $2\pi x_{\perp}$ and $4\pi r^{2}$ respectively. one can see from Fig. \ref{massdistributions} that 3D mass distribution exhibits a broader shape than the 2D mass distributions. This indicates that the 3D mass-radius \cite{Goeke:2007fp}  should be larger than the 2D mass-radius \cite{Kim:2021jjf}. The numerical values of the mass radii for the 2D and 3D distributions are given in Table \ref{table2}. The  ratio between the 2D and 3D mean square  mass radii in this model is found as 
	\begin{eqnarray}
		\frac{\langle x_{\perp}^{2}\rangle_{mass}}{\langle r^{2}\rangle_{mass}}=\frac{2}{3},
	\end{eqnarray}
	where these 2D and 3D mass radii are respectively defined as \cite{Freese:2021czn,Goeke:2007fp,Kim:2021jjf}
	\begin{eqnarray}
		\langle x_{\perp}^{2}\rangle_{mass}=\frac{1}{M}\int d^{2}x_{\perp} x_{\perp}^{2} \mathcal{E}(x_{\perp}), \hspace*{1cm} \langle r^{2}\rangle_{mass}=\frac{\int d^{3}r r^{2}\epsilon(r)}{\int d^{3}r \epsilon(r)}.
	\end{eqnarray}
	
	\begin{figure}
		\includegraphics[scale=0.38]{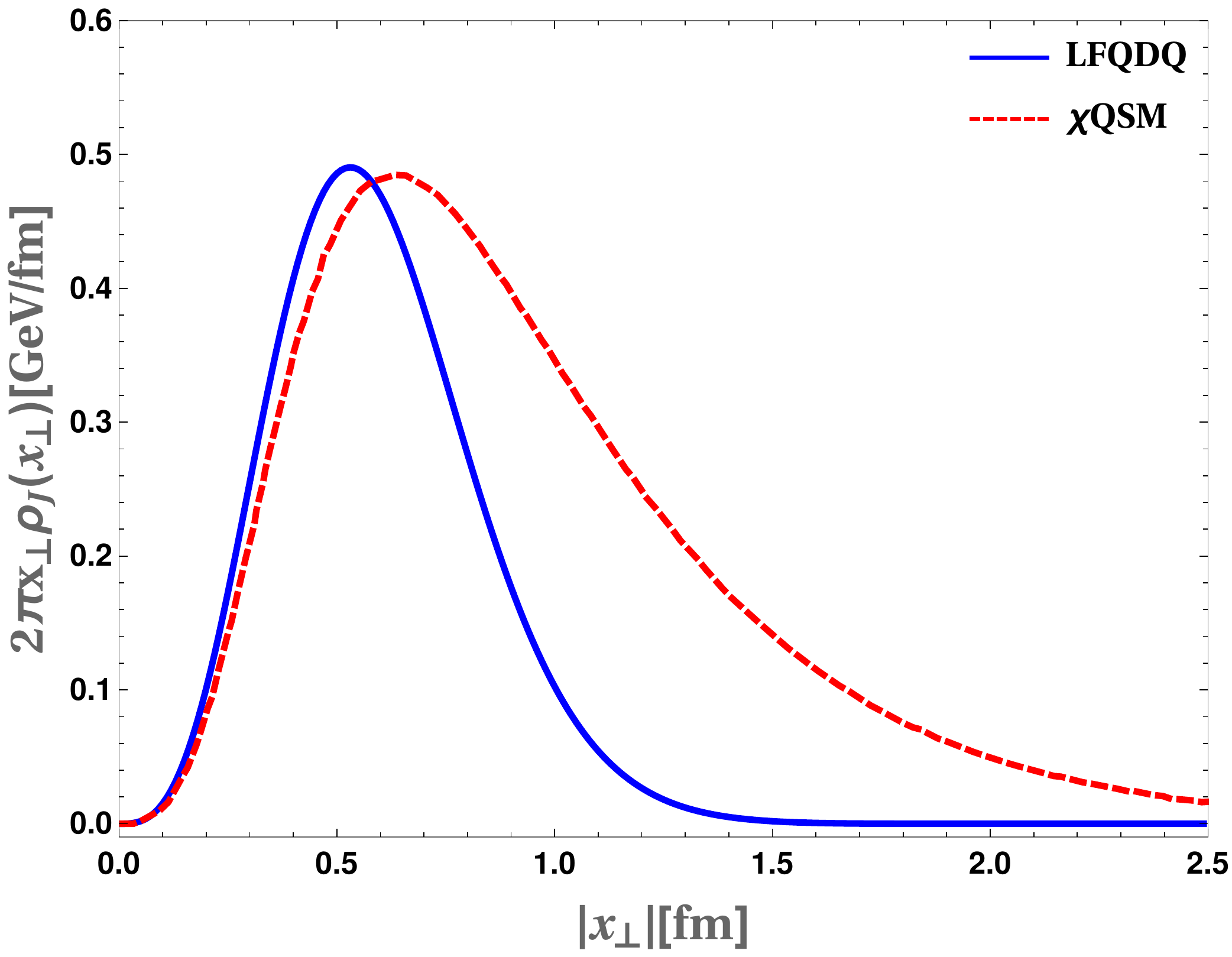}
		\includegraphics[scale=0.38]{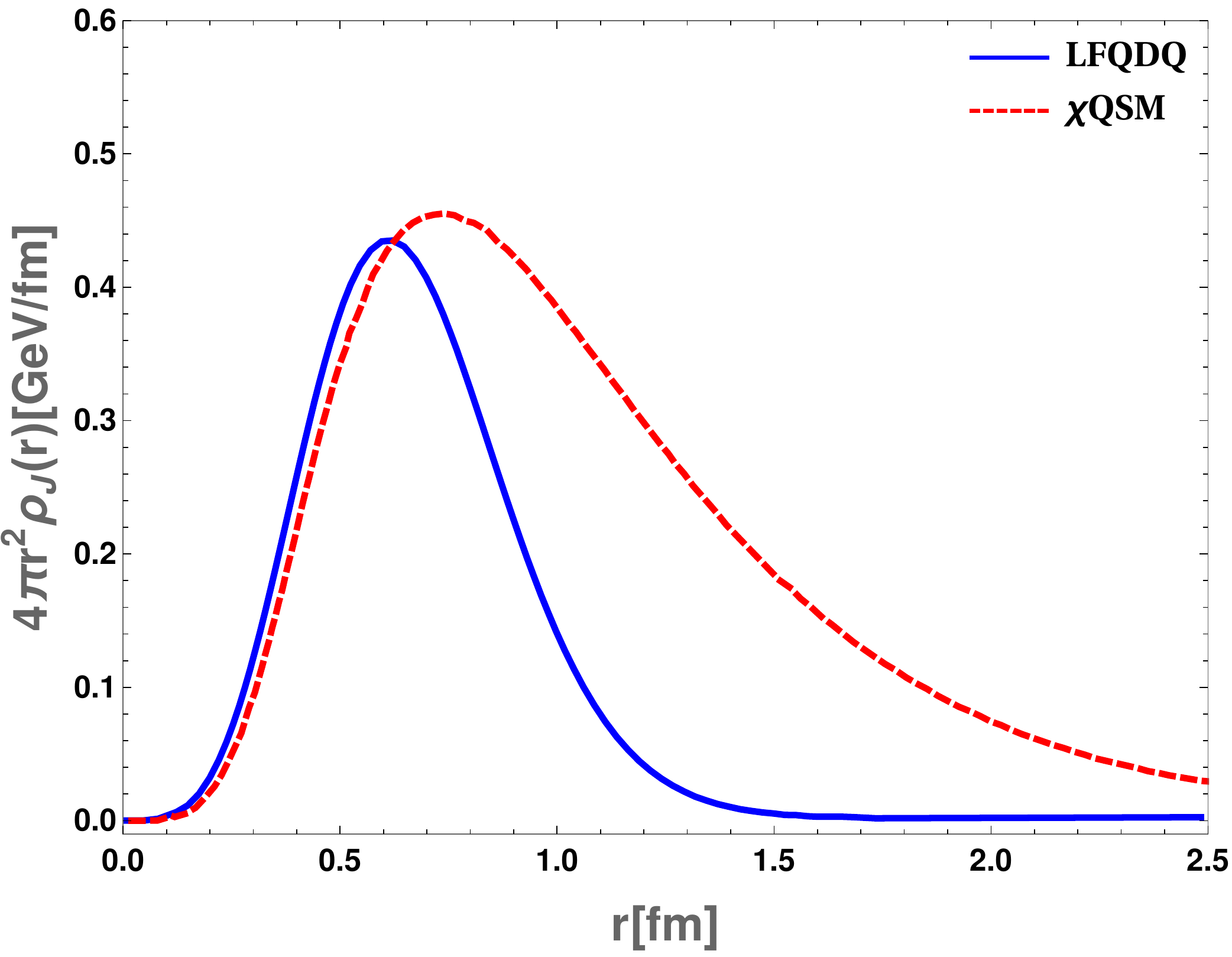}
		\caption{The solid blue curve in the left panel shows the 2D angular momentum distribution in the LF-frame for the LFQDQ model while the red-dashed curve is the 2D angular momentum distribution for the $\chi$QSM model \cite{Kim:2021jjf}. Similarly, the solid blue curve in the right panel shows the 3D angular momentum distribution in the BF for the LFQDQ (after doing the inverse Abel transformation) whereas the red-dashed curve is the 3D angular momentum distribution for $\chi$QSM model \cite{Goeke:2007fp}. Our model predictions are at the  scale $\mu^{2}$=4 GeV$^{2}$.}
		\label{AngularMom3D}
	\end{figure}
	\begin{figure}
		\includegraphics[scale=0.38]{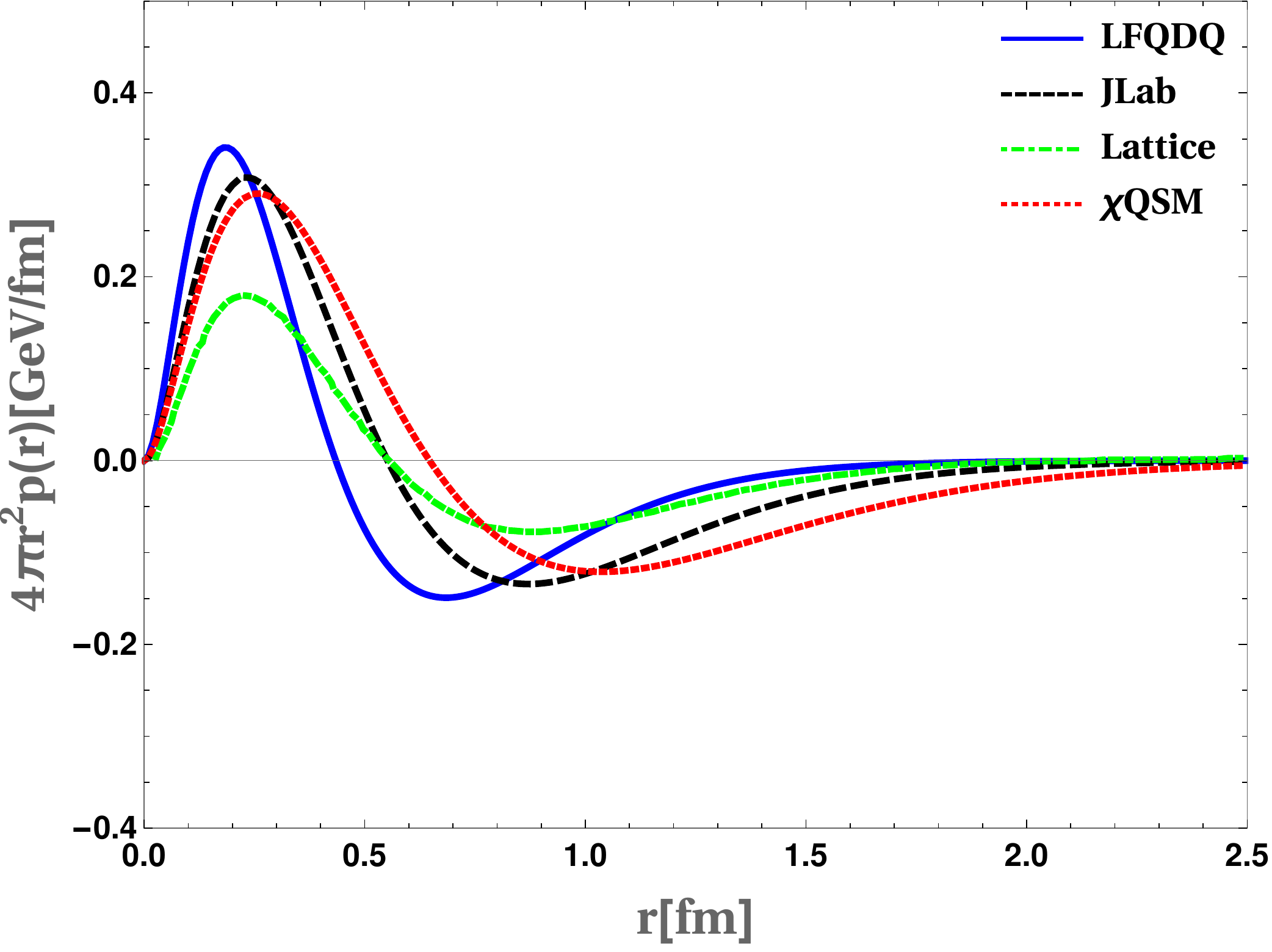}
		\includegraphics[scale=0.38]{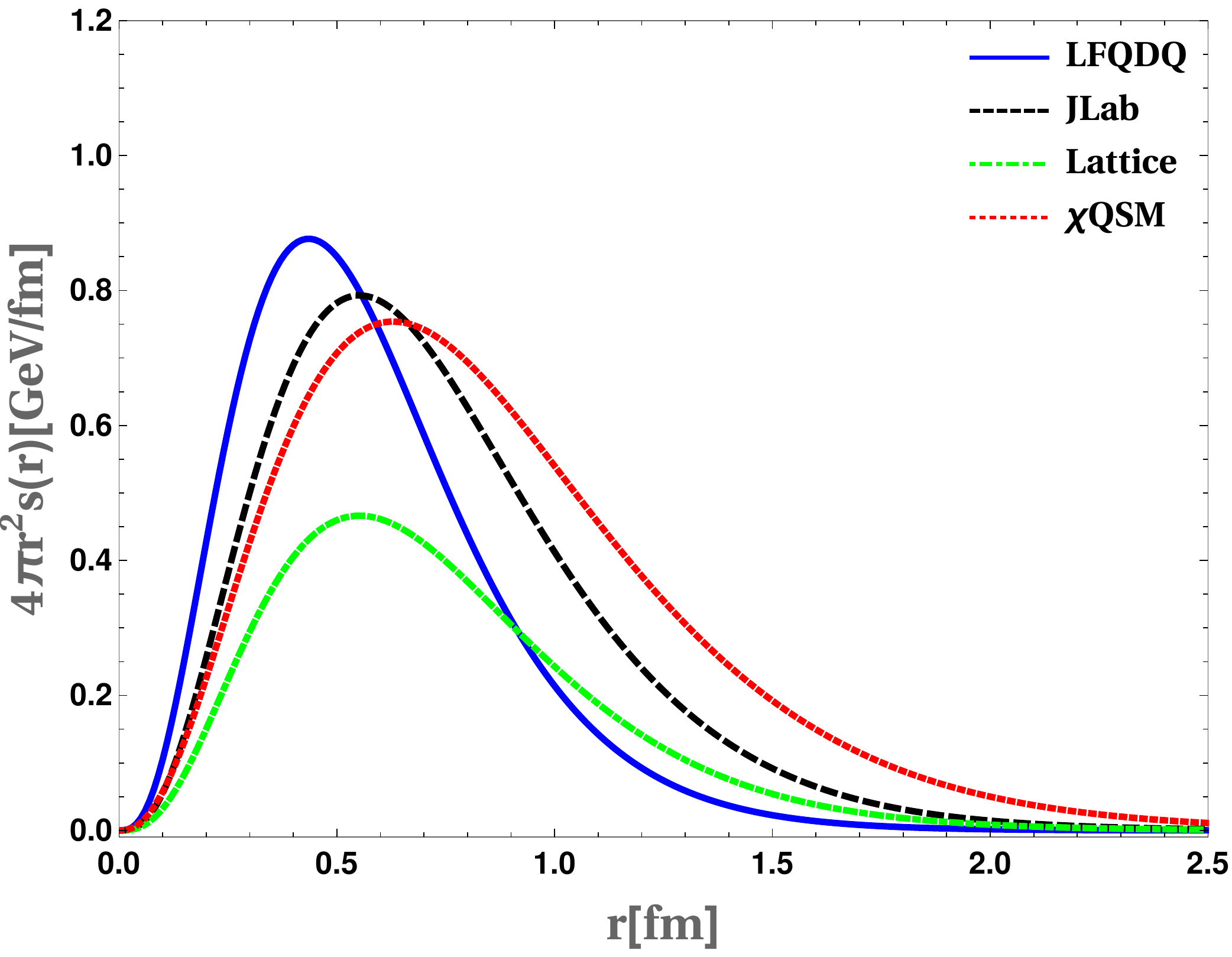}
		\caption{The solid blue, black-dashed, red-dotted, and green-dot-dashed curves in the left panel represent the 3D pressure distributions (in Breit Frame) for LFQDQ, JLab \cite{Burkert:2018bqq,Kumericki:2019ddg}, $\chi$QSM \cite{Goeke:2007fp} and Lattice predictions \cite{Shanahan:2018nnv}, respectively. Whereas Solid blue, black-dashed, red-dotted, and green-dot-dashed curves in the right panel show the 3D shear-force distributions in the BF for LFQDQ, JLab \cite{Burkert:2018bqq,Kumericki:2019ddg}, $\chi$QSM \cite{Goeke:2007fp} and Lattice predictions \cite{Shanahan:2018nnv}. Our model predictions are at the  scale $\mu^{2}$=4 GeV$^{2}$.}
		\label{3D-pressure-shear}
	\end{figure}
	In Fig. \ref{AngularMom3D} the solid-blue  curves in the left panel represent the 2D angular momentum distribution (weighted by $2\pi x_{\perp}$)  calculated by using  Eq.(\ref{2Denergy-angular momentum}),  for the LFQDQ  model whereas the dashed-red curves represent the same in  $\chi$QSM model. The solid-blue (dashed-red) curves in the right panel show the 3D angular momentum distribution weighted by $4\pi r^{2}$ which are computed by using the inverse Abel transform  given in Eq.(\ref{3DAngularMom}), in the LFQDQ ($\chi QSM$) model. The 2D and 3D angular-momentum distributions are normalized as 
	\begin{eqnarray}
		\int d^{2}x_{\perp}\rho^{2D}_{J}(x_{\perp})=\int d^{3}r \rho_{J}(r)=J(0)=\frac{1}{2},
	\end{eqnarray}
	which is related to the nucleon spin. Similar to the mass distribution, the 3D angular momentum distribution is also broader than the 2D distribution. The 2D radius \cite{Kim:2021jjf} for the angular momentum distribution is related to the 3D distribution  \cite{Goeke:2007fp}, and is smaller than the 3D radius by a geometric factor 4/5. i.e,
	\begin{eqnarray}
		\langle x_{\perp}^{2}\rangle_{J}\approx \frac{4}{5}\langle r^{2}\rangle_{J}
	\end{eqnarray}
	where these 2D and 3D angular momentum radii are defined as 
	\begin{eqnarray}
		\langle x_{\perp}^{2}\rangle_{J}=2 \int d^{2}x_{\perp}x_{\perp}^{2}\rho_{J}^{(2D)}(x_{\perp}), \hspace*{0.5cm} and \hspace*{0.5cm} \langle r^{2}\rangle_{J}=\frac{\int d^{3}r r^{2}\rho_{J}(r)}{\int d^{3}r \rho_{J}(r)}.
	\end{eqnarray}
	The numerical values of these 2D and 3D angular momentum radii for the LFQDQ model are given in Table \ref{table2}.
	
	The solid-blue curve in the left(right) panel of Fig. \ref{3D-pressure-shear} shows the 3D pressure(shear-force) distributions for the LFQDQ model, while the black-dashed, red-dotted and green-dot-dashed curves show the pressure and shear-force distributions for JLab, $\chi QSM$ and Lattice predictions, respectively. The pressure $p(r)$ has the global maximum at $r=0$, with $p(0)=4.76 GeV/fm^{3} =7.62 \times 10^{35}$ Pa. Which is 10-100 times higher than the pressure inside a neutron star \cite{Prakash:2000jr}. The pressure decreases monotonically, becoming zero at the nodal point, $r_{0}\approx0.43$ fm. The pressure reaches the global minimum at $r_{p,min}=0.67$ fm, after which it increases monotonically but remains  negative until it goes to zero. The positive sign of the pressure for $r<r_{0}$ corresponds to the repulsion, whereas the negative sign in the region $r>r_{0}$ is for the attraction. 
	Unlike  pressure, the shear force  distribution is always positive.   
	
	The conservation of the EMT currents $\partial_{\mu}\hat{T}^{\mu\nu}=0$, provides the 2D as well as 3D stability conditions. We obtain the 3D equilibrium equations from the conservation of the EMT currents, which are equivalent to the 2D stable conditions as \cite{Polyakov:2018zvc,Kim:2021jjf,Panteleeva:2021iip},
	\begin{eqnarray}\label{2D-3D stability}
		p'(r)+\frac{2s(r)}{r}+\frac{2}{3}s'(r)=0 \Longleftrightarrow \mathcal{P}'(x_\perp)+\frac{\mathcal{S}(x_{\perp})}{x_{\perp}}+\frac{1}{2}\mathcal{S}'(x_\perp)=0
	\end{eqnarray}
	From the above Eq.(\ref{2D-3D stability}) one can easily see that the pressure and shear forces are not independent functions but due to EMT conservation, they are related to each other. Another consequence of the EMT conservation is the von Laue condition for the 2D and 3D pressure and shear forces for the nucleons \cite{Polyakov:2018zvc},
	\begin{eqnarray}\label{von-laue}
		\int d^{3}rp(r)=0 \Longleftrightarrow \int d^{2} x_{\perp} \mathcal{P}(x_{\perp})=0
	\end{eqnarray}
	\begin{eqnarray}\label{stability2}
		\int_{0}^{\infty} dr r \left[p(r)-\frac{1}{3}s(r)\right]=0\Longleftrightarrow \int_{0}^{\infty} dx_{\perp}\left[\mathcal{P}(x_{\perp})-\frac{1}{2}\mathcal{S}(x_\perp)\right]=0
	\end{eqnarray}
	From the above two Eqs. ((\ref{von-laue}),(\ref{stability2})) one can see that 3D von Laue conditions are satisfied if and only if the 2D ones are satisfied. By using the Eq.(\ref{2D-3D stability}) the Druck-term can be expressed in terms of 2D and 3D pressure and force distributions as,
	\begin{eqnarray}\label{D(0)fromp(x)}
		D(0)=-M\int d^{2}x_{\perp}x_{\perp}^{2}\mathcal{S}(x_\perp)=4M\int d^{2}x_{\perp}x_{\perp}^{2}\mathcal{P}(x_\perp) 
	\end{eqnarray}
	and,
	\begin{eqnarray}\label{D(0)fromp(r)}
		D(0)=-\frac{4}{15}M\int d^{3}r r^{2}s(r)=M\int d^{3}r r^{2}p(r)
	\end{eqnarray}
	respectively. It indicates that the 3D Able images of the 2D distributions show the equivalently same mechanical properties as 2D distributions.
	
	In Refs. \cite{Lorce:2018egm,Polyakov:2018zvc,Perevalova:2016dln} it was shown that for the local stability of the mechanical system, the 3D and 2D pressure and shear forces should satisfy the following conditions
	\begin{eqnarray}\label{localstability}
		\frac{2}{3}s(r)+p(r)>0, \hspace*{1cm} {\rm and} \hspace*{1cm}\frac{1}{2}\mathcal{S}(x_{\perp})+\mathcal{P}(x_{\perp})>0.
	\end{eqnarray}
	This inequalities imply that the Druck term (D-term) for any stable system must be negative, i.e., $D(0)<0$.
	%From Eq.(\ref{localstability}) one can see that the Able image of a positive function is also positive. 
	More discussions about the local stability (Eq.(\ref{localstability})) can be found in  Ref.\cite{Freese:2021czn}, it is an interesting result that the stability condition in 3D implies the stability of the 2D mechanical system \cite{Chakrabarti:2020kdc}. This allows us to connect the 3D mechanical radius to that in 2D  as
	\begin{eqnarray}
		\langle x_{\perp}^{2}\rangle_{mech}=\frac{4D(0)}{\int_{-\infty}^{0} dt D(t)}=\frac{2}{3}\langle r^{2}\rangle_{mech}
	\end{eqnarray}
	where the 2D mechanical radius is defined as 
	\begin{eqnarray}
		\langle x_{\perp}^{2}\rangle_{mech}=\frac{\int d^{2}x_{\perp}x^{2}_{\perp}\left(\frac{1}{2}\mathcal{S}(x_\perp)+\mathcal{P}(x_\perp)\right)}{\int d^{2}x_{\perp}\left(\frac{1}{2}\mathcal{S}(x_\perp)+\mathcal{P}(x_\perp)\right)}
	\end{eqnarray}
	and the mechanical radius in 3D is given by
	\begin{eqnarray}
		\langle r^{2}\rangle_{mech}=\frac{\int d^{3}r r^{2}\left(\frac{2}{3}s(r)+p(r)\right)}{\int d^{3}r\left(\frac{2}{3}s(r)+p(r)\right)}.
	\end{eqnarray}
	\begin{figure}
		\includegraphics[scale=0.385]{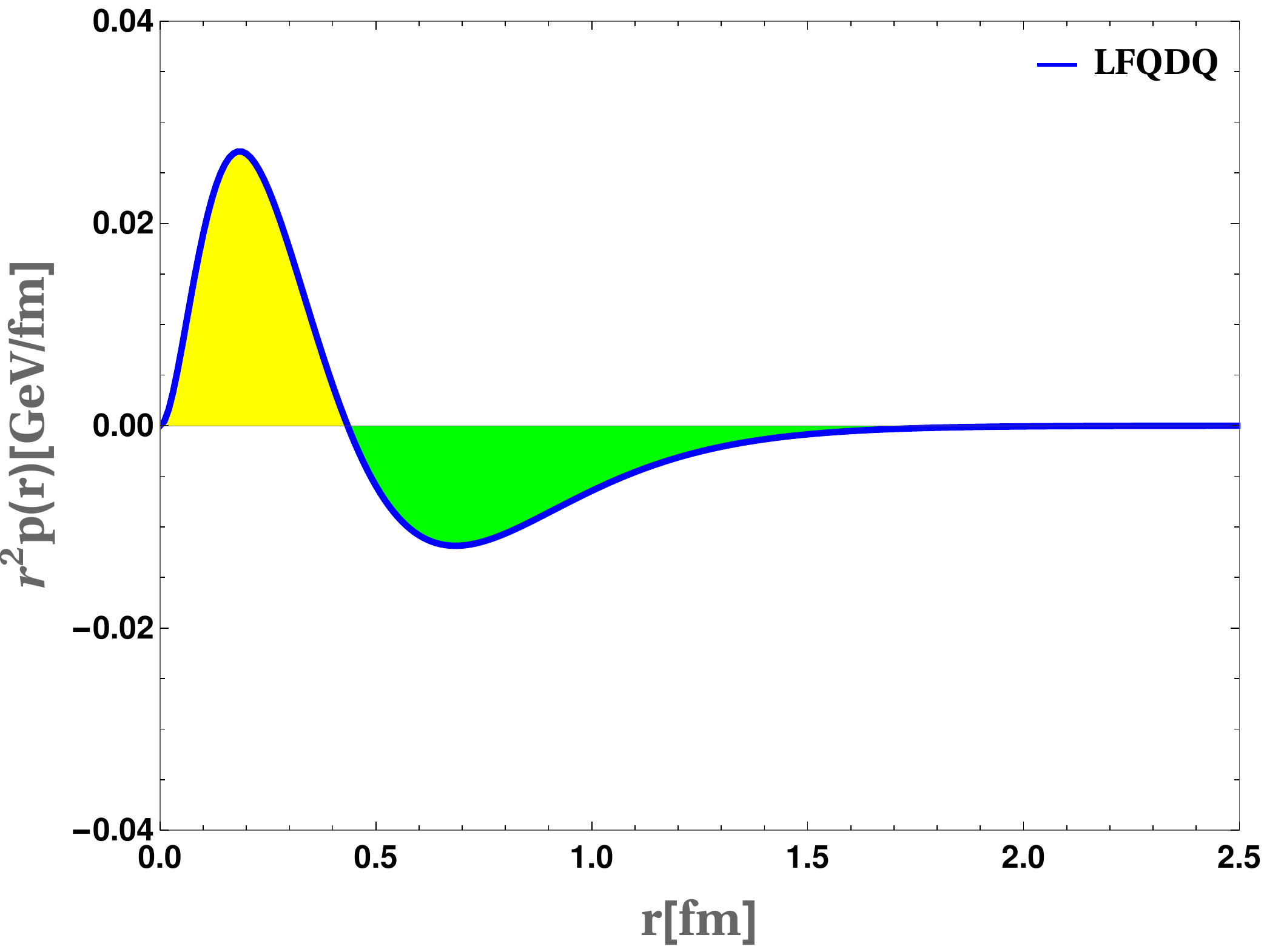}
		\includegraphics[scale=0.38]{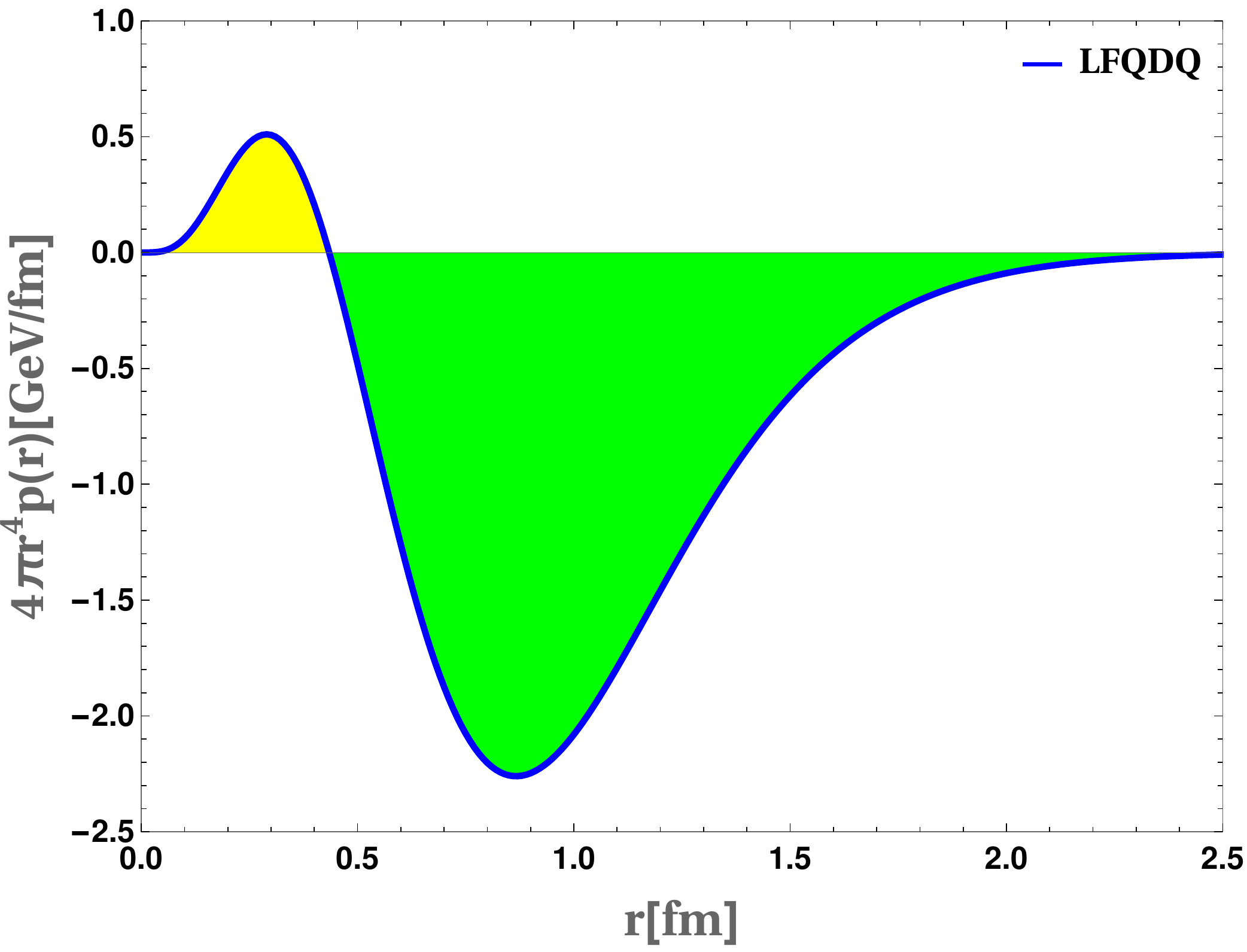}
		\caption{Left panel shows $r^{2}p(r)$ as a function of r from the LFQDQ model at evolved scale $\mu^{2}$=4 GeV$^{2}$. It shows  how the stability condition $\int_{0}^{\infty}drr^{2}p(0)=0$ in Eq.(\ref{von-laue}) is realized.  Right panel shows $4\pi r^{4}p(r)$. Note that the area under the curve is negative which implies $D<0$.}
		%gives the D-term as Eq.(\ref{D(0)fromp(r)}).}
	\label{3Dstability}
\end{figure}
%are the 2D and 3D mechanical radii, respectively.
Numerical verification of  the stability condition Eq.(\ref{von-laue}) is presented in Fig. \ref{3Dstability} . The left panel in Fig. \ref{3Dstability} shows $r^{2}p(r)$ as a function of $r$. The yellow shaded region in the positive upper half in the left panel of Fig. \ref{3Dstability}  has exactly the same surface areas as in the negative half (shaded in green). i.e.,
\begin{eqnarray}
	\int_{0}^{r_{0}}dr r^{2}p(r)=6.74 MeV,\\ \nonumber
	\int_{r_0}^{\infty}dr r^{2}p(r)=-6.74 MeV
\end{eqnarray}
\begin{table}
	$$
	\begin{array}{lcccccc}
		\hline \hline \mathcal{E}(0)\left(\mathrm{GeV} / \mathrm{fm}^{2}\right) & \mathcal{P}(0)\left(\mathrm{GeV} / \mathrm{fm}^{2}\right) & \left(x_{\perp}\right)_{0}(\mathrm{fm}) & \left\langle x_{\perp}^{2}\right\rangle_{\text {mass }}\left(\mathrm{fm}^{2}\right) & \left\langle x_{\perp}^{2}\right\rangle_{J}\left(\mathrm{fm}^{2}\right) & \left\langle x_{\perp}^{2}\right\rangle_{\text {mech }}\left(\mathrm{fm}^{2}\right) \\
		\hline 1.54 & 0.354 & 0.34 & 0.21 & 0.38 & 0.167  \\
		\hline \hline \varepsilon(0)\left(\mathrm{GeV} / \mathrm{fm}^{3}\right) & p(0)\left(\mathrm{GeV} / \mathrm{fm}^{3}\right) & r_{0}(\mathrm{fm}) & \left\langle r^{2}\right\rangle_{\text {mass }}\left(\mathrm{fm}^{2}\right) & \left\langle r^{2}\right\rangle_{J}\left(\mathrm{fm}^{2}\right) & \left\langle r^{2}\right\rangle_{\text {mech }}\left(\mathrm{fm}^{2}\right)  \\
		\hline 2.02 & 4.76 & 0.43 & 0.32 & 0.51 & 0.251  \\
		\hline \hline
	\end{array}
	$$
	\caption{Various observable obtained from the EMT distributions for the proton in both  2D LF and 3D BF are listed: the energy distributions at the nucleon center ($\mathcal{E}(0)$,$\epsilon(0)$), pressure distribution at the nucleon center ($\mathcal{P}(0)$,$p(0)$), nodal points of the pressure ($(x_{\perp})_{0}$,$r_{0}$), and the mean square radii of the mass, angular momentum and mechanical ($\langle x_{\perp}^{2}\rangle$, $\langle r^{2}\rangle$).}
	\label{table2}
\end{table}
where $r_{0}$ is the nodal point in 3D BF, and thus they cancel each other to produce zero as required by the stability condition (Eq.(\ref{von-laue})).
In the right panel of Fig. \ref{3Dstability} we show $4\pi r^{4}p(r)$ with $r$ which tells us about the sign of the D-term. The area in the negative half (green) is much larger than the area in the positive half (yellow).  From Eq.(\ref{D(0)fromp(r)}) we can see that in the LFQDQ model the D-term at zero momentum transfer takes a negative value, i.e, $D(0)<0$. The same conclusion can be derived from Eq.(\ref{D(0)fromp(x)}) as well.

The pressure and the shear force distributions are again related to %considered as the notion of 
the normal(radial) and the tangential force fields, which are the eigenvalues of the stress tensor, $T_{ij}$. %with $e_{r}$ and $e_{\phi}$ unit vectors. 
So, the 3D and the 2D force fields on the BF and the LF frame can be obtained as \cite{Polyakov:2018zvc,Kim:2021jjf},
\begin{eqnarray}\label{Fn3D-Ft3D}
	F_{n}(r)=4\pi r^{2}\left[\frac{2}{3}s(r)+p(r)\right], \hspace*{1cm} F_{t}(r)=4\pi r^{2}\left[-\frac{1}{3}s(r)+p(r)\right]
\end{eqnarray}  
and,
\begin{eqnarray}\label{Fn2D-Ft2D}
	F_{n}^{(2D)}(x_{\perp})=2\pi x_{\perp}\left[\frac{1}{2}\mathcal{S}(x_{\perp})+\mathcal{P}(x_\perp)\right], \hspace*{1cm} F_{t}^{(2D)}(x_{\perp})=2\pi x_{\perp}\left[-\frac{1}{2}\mathcal{S}(x_{\perp})+\mathcal{P}(x_\perp)\right]
\end{eqnarray}
\begin{figure}
	\includegraphics[scale=0.38]{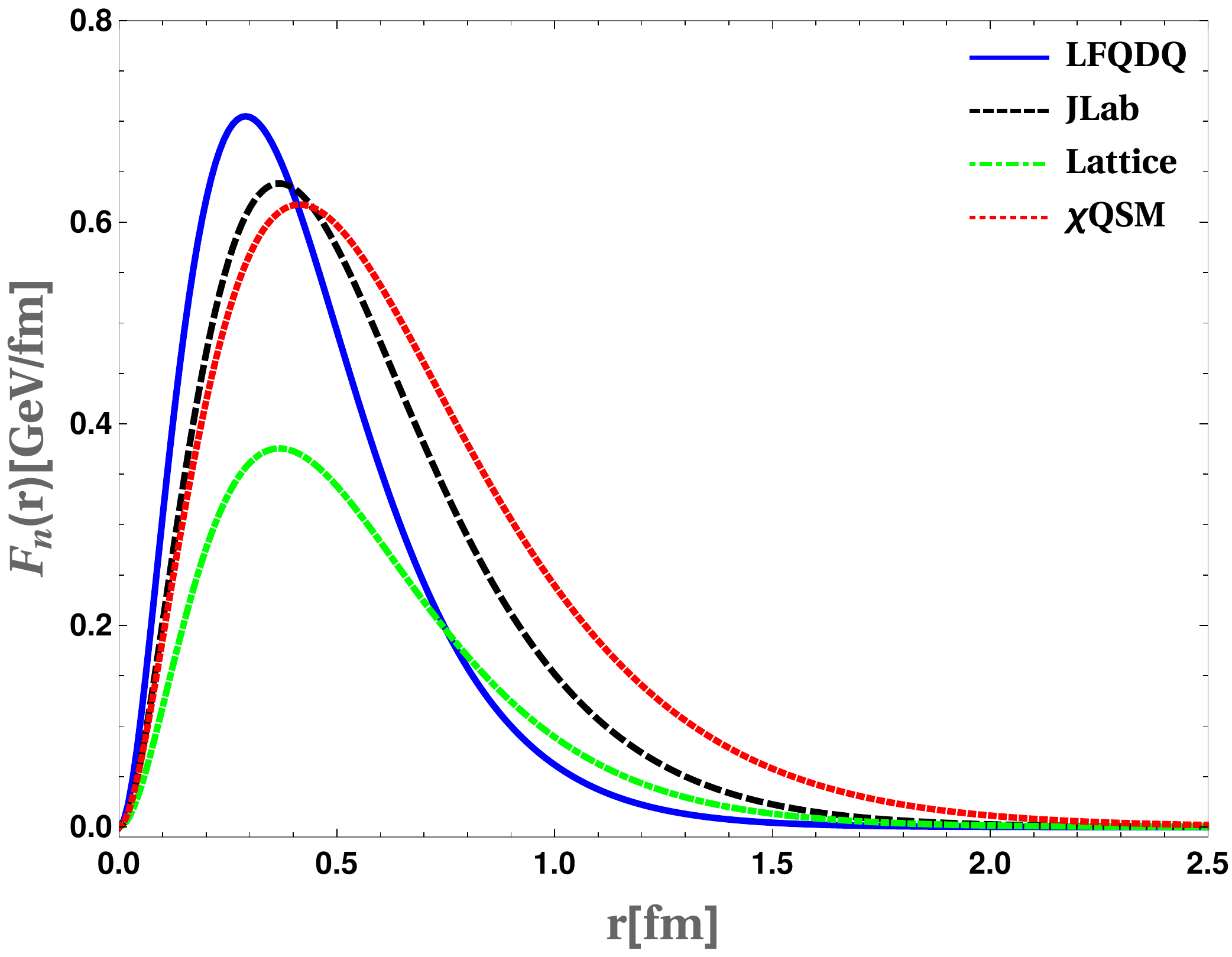}
	\includegraphics[scale=0.38]{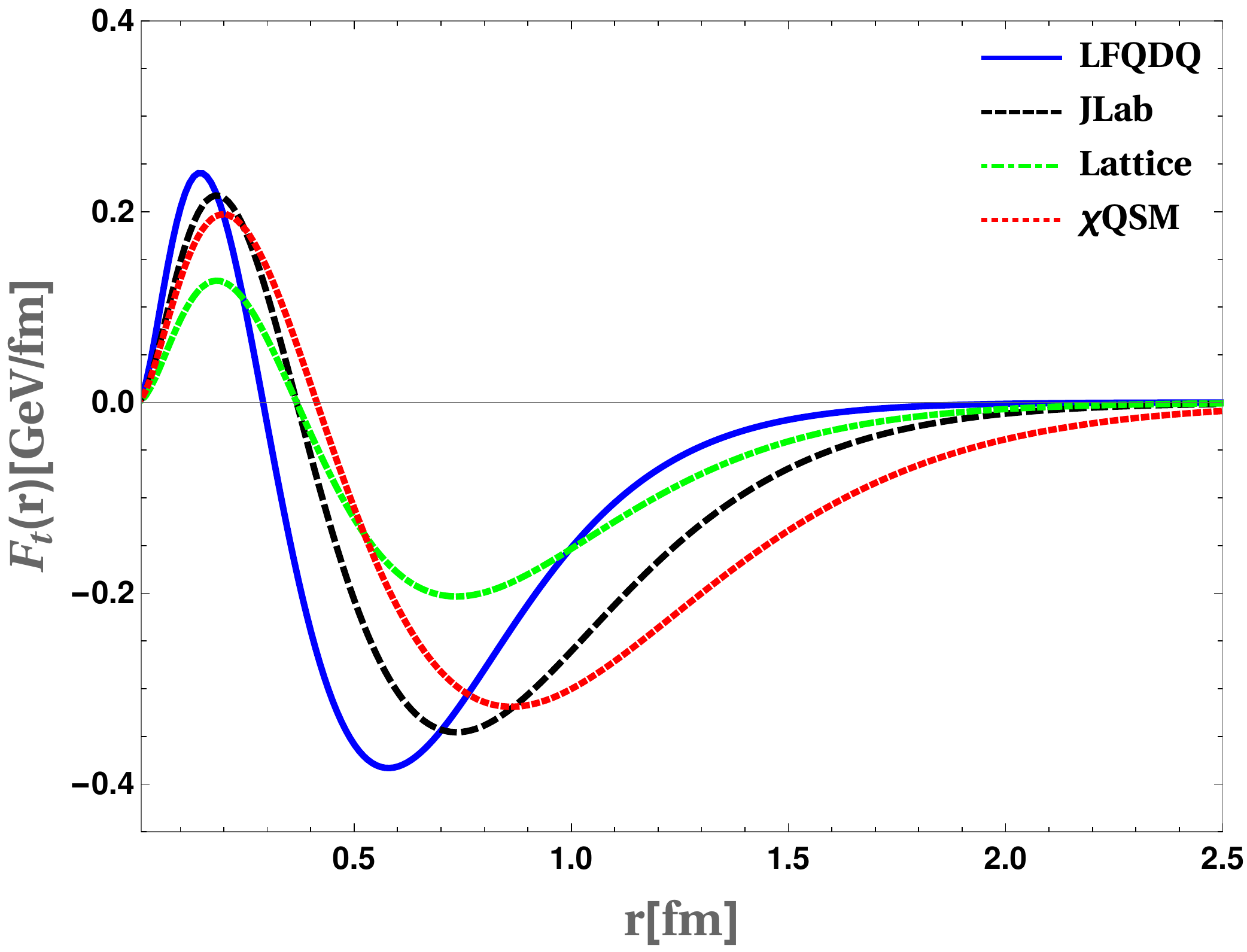}
	\caption{The solid blue, black-dashed, red-dotted and green-dot-dashed curves in the left panel represents the 3D normal force field distributions (in Breit Frame) for LFQDQ, JLab \cite{Burkert:2018bqq,Kumericki:2019ddg}, $\chi$QSM \cite{Goeke:2007fp} and Lattice predictions \cite{Shanahan:2018nnv}, respectively. Whereas Solid blue, black-dashed, red-dotted, and green-dot-dashed curves in the right panel show the 3D tangential force field distributions in the BF for LFQDQ, JLab \cite{Burkert:2018bqq,Kumericki:2019ddg}, $\chi$QSM \cite{Goeke:2007fp} and Lattice predictions \cite{Shanahan:2018nnv}, respectively. Our model predictions are at the evolved scale $\mu^{2}$=4 GeV$^{2}$.}
	\label{3D-Fn-Ft}
\end{figure}
respectively, In the left panel of Fig. \ref{3D-Fn-Ft} the solid blue curve depicts the 3D normal force field, while the black-dashed, red-dotted and the green-dot-dashed curves are the 3D normal force distributions for the JLab \cite{Burkert:2018bqq,Kumericki:2019ddg}, $\chi$QSM \cite{Goeke:2007fp} and Lattice predictions \cite{Shanahan:2018nnv}, respectively. The right panel of Fig. \ref{3D-Fn-Ft}  shows the 3D tangential force field distributions for the LFQDQ (solid-blue), JLab (black-dashed) \cite{Burkert:2018bqq,Kumericki:2019ddg}, $\chi QSM$ (red-dotted) \cite{Goeke:2007fp} and Lattice (green-dot-dashed) \cite{Shanahan:2018nnv}, respectively. The 2D normal and tangential force field distributions for the quark- scalar-diquark model can be found in  Ref.\cite{Chakrabarti:2020kdc}. In a stable spherically symmetric system the normal force $F_{n}(r)$ must be a stretching force otherwise the system would squeeze and collapse to the center. Whereas, the tangential force changes its direction with the distance $r$ because the average value of possible squeezing has to be zero for a spherically symmetric system. The normal force complies with the local stability condition (\ref{localstability}) and the tangential force satisfies the von Laue condition
(\ref{von-laue}). Due to this condition, both 2D and 3D tangential forces have at least one nodal point, which tells that the direction of the force field should be reversed at this point. In the LFQDQ model the 3D and 2D tangential forces changes its direction at $r_{0}\approx0.29$ fm and $(x_{\perp})_{0}\approx 0.20$ fm, respectively.

In Table \ref{table2}, we list the numerical values for various observables such as energy and pressure densities at the center of the nucleon  in both the BF and LF frames. The explicit values of the nodal points are also given. One can see from Table \ref{table2} that the magnitudes of these observables are larger in 3D BF than those in the 2D LF frame. A similar type of behavior for those observables  has been also observed  in  Ref.\cite{Kim:2021jjf}.
%%%%%%%%%%%%%%%%%%%%%%%%%%%%%%%%%%%%%%%%%%%%%%%%%%%%%%%%%%%%%%%%%%%%%%%%%%%%%%%%%%%%%%%%%%%%%%%%%%%%%%% 

\section{Conclusion}\label{conclusion}	
Of the three GFFs, Druck term or the D-term is the least understood form factor. The D-term is physically very important as it gives the shear and pressure distributions inside the proton. Recently, JLab reported the first measurement of the shear and pressure forces inside the proton and hence there are renewed interests in recent time to study the D-term in different models. Generally, the three dimensional distributions are defined in the Breit frame which are subject to relativistic corrections while in the light front the distributions are most conveniently evaluated in 2D transverse plane. Recently,  it was shown that the Abel transformation relates the 2D light front distributions to the 3D distributions in the Breit frame. In this paper, the 2D LF distributions are evaluated in a quark-scalar diquark model of proton and then the 3D distributions are obtained using the Abel transformation. The wavefunctions in the model are constructed by modifying the two-particle wavefunctions predicted by AdS/QCD which can not be evaluated in perturbation theory and encode nonperturbative contributions.  Our results are compared with the $\chi QSM$, JLab and lattice predictions. The 3D stability conditions translated to 2D  are found to be satisfied by the 2D distributions obtained in the LFQDQ model. Various properties such as the energy and pressure distributions at the nucleon center, mass, angular momentum, mechanical radii, etc are evaluated from the EMT distributions in 2D transverse plane and the corresponding 3D distributions in the Breit frame are obtained by Abel transformations.  The normal and shear force distributions are also evaluated in the LFQDQ model. Our results are found to be consistent with lattice and other model predictions. 

\section{Acknowledgement}

This work is  supported by Science and Engineering Research Board under the
Grant No. CRG/2019/000895. The work of AM is supported by the SERB-POWER Fellowship (file no. SPF/2021/000102).

\bibliographystyle{unsrt}
\bibliography{References}
%\bibliography{ref}
\end{document}